\documentclass[journal]{IEEEtran}
\usepackage{cite}
\usepackage{multirow}
\usepackage{mathrsfs}
\usepackage{graphicx}
\usepackage{epstopdf}
\usepackage{amsfonts}
\usepackage{amsmath,amssymb}
\usepackage{mathrsfs}
\usepackage{pifont}
\usepackage{bbding}
\usepackage{amssymb}
\usepackage{subfigure}
\usepackage{color}
\usepackage[justification=centering]{caption}
\usepackage{subeqnarray}
\usepackage{cases}
\usepackage{float}
\usepackage{bm}
\usepackage{multicol}

\newtheorem{theorem}{Theorem}
\newtheorem{lemma}{Lemma}
\newtheorem{corollary}{Corollary}
\newtheorem{remark}{Remark}

\begin{document}
\title{Tracking Performance Limitations of MIMO Networked Control Systems with Multiple Communication Constraints}
\author{Chao-Yang Chen, Weihua Gui, Lianghong Wu, Zhaohua Liu, Huaicheng Yan
\renewcommand{\thefootnote}{\fnsymbol{footnote}}
\thanks{This work is partially supported by the Foundation for the National Natural Science Foundation of China (61503133,61673178), Innovative Research Groups of the National Natural Science Foundation of China (61321003), and Natural Science Foundation of Hunan (2016JJ6043,2018JJ2137,2018JJ2134), Hunan Provincial young talents project (2018RS3095), Shanghai Natural Science Foundation (17ZR1444700), Shanghai Shuguang Project (16SG28), Program of Shanghai Academic Research Leader (19XD1421000). (\emph{Corresponding author: Chao-Yang Chen})}
\thanks{C. -Y. Chen is with School of Information and Electrical Engineering, Hunan University of Science and Technology, Xiangtan 411201, China, and with School of Information Science and Engineering, Central South University, Changsha 410012, China, and also with Center for Polymer Studies and Department of Physics, Boston University, Boston, MA 02215, USA (Email: ouzk@163.com.)}
\thanks{W. Gui is with School of Information Science and Engineering, Central South University, Changsha 410012, China (Email: gwh@mail.csu.edu.cn).}
\thanks{L. Wu and Z. Liu are with School of Information and Electrical Engineering, Hunan University of Science and Technology, Xiangtan 411201, China (Email: lhwu@hnust.edu.cn; zhaohualiu2009@hotmail.com).}
\thanks{H. Yan is with the Key Laboratory of Advanced Control and Optimization for Chemical Process of Ministry of Education, East China University of Science and Technology, Shanghai 200237, China, and also with the College of Mechatronics and Control Engineering, Hubei Normal University, Huangshi 435002, China (e-mail: hcyan@ecust.edu.cn).}}
\date{}
\maketitle
\hspace{-0.5cm}
\begin{abstract}
In this paper, the tracking performance limitation of networked control systems (NCSs) is studied. The NCSs is considered as continuous-time linear multi-input multi-output (MIMO) systems with random reference noises. The controlled plants include unstable poles and non-minimum phase (NMP) zeros. The output feedback path is affected by multiple communication constraints. We focus on some basic communication constraints, including additive white noise (AWN), quantization noise, bandwidth, as well as encoder-decoder. The system performance is evaluated with the tracking error energy, and used a two-degree of freedom (2DOF) controller. The explicit representation of the tracking performance is given in this paper. The results indicate the tracking performance limitations rely to internal characteristics of the plant (unstable poles and NMP zeros), reference noises (the reference noise power distribution (RNPD) and its directions) and the characteristics of communication constraints. The characteristics of communication constraints include communication noise power distribution (CNPD), quantization noise power distribution (QNPD), and their distribution directions, transform bandwidth allocation (TBA), transform encoder-decoder allocation (TEA), and their allocation directions, and NMP zeros and MP part of bandwidth. Moreover, the tracking performance limitations are also affected by the angles between the each transform NMP zero direction and RNPD direction, and these angles between each transform unstable poles direction and the direction of communication constraint distribution/allocation. In addition, for MIMO NCSs, bandwidth (there are not identical two channels) always can affects the direction of unstable poles, and the channel allocation of bandwidth and encode-decode may be used for a feasible method for the performance allocation of each channels. Lastly, a instance is given for verifying the effectiveness of the theoretical outcomes.
\end{abstract}


\section{INTRODUCTION}\label{se1}
\IEEEPARstart{I}{n} the past ten years, there has been growing concern about NCSs. It is well accepted that NCSs take both aspects, control and communication, into consideration simultaneously \cite{Wen17,Hu18,17Zhang,17WuAsynchronous,chen2017adaptive,ge2016task,wang2018optimal,zheng2019active}. They have been applied in many important areas, e.g., industry control applications, automobiles, factory automation, intelligent traffic, etc. Nowadays, the main issues addressed include modeling and stabilization analysis of NCSs with communication restrictions, including data-rate constraints\cite{16Matveev,07Braslavsky}, bandwidth restrictions\cite{13Guan,zhan2019optimal}, time-delays\cite{yan2017novel,yany2019input}, quantization\cite{chen2017fundamentalb,16NingAsynchronous}, packet losses\cite{12XuStochastic,15LiWu} as well as communication noises\cite{fang2017design,zhang2018adaptive}. Notwithstanding the spectacular achievements of these studies, even more thought-provoking cases of best accomplishable tracking capability in certain network condition still require to be further analyzed. The analysis of the control system mainly includes two aspects: stability and performance \cite{seron1997fundamental}. The NCSs as a special type of control systems, its optimal performance naturally becomes a new challenge, and it is crucial to analyzing how communication constraints affect its performance. This will inspire the design of the NCSs.
\par
Performance limitations as an important branch of control system performance research\cite{seron1997fundamental,chen2003guest}, it has a lot of results in the control community\cite{seron1997fundamental,chen2000limitations,freudenberg2003fundamental,middleton2004tracking}. The essential feature of control systems is the intrinsic accomplishable performance limitations, which remains a constant no matter what the controller is adopted. The limitation is dependent on the inner characteristics of the plant (like NMP zeros and unstable poles). For classical control systems, an exact expression of performance limitations has been obtained in \cite{chen2000limitations}. After that, there are also many meaningful extended works, e.g., Wang et al. \cite{11Wang} analyzes the tracking capability of linear time-invariant (LTI) systems. It is shown that the properties of the plant are affected by NMP zeros, unstable poles, their directions. In addition, it has shown that the 2DOF controller would bring improvement of the performance in the control systems. However, the typical best control capability is attained as idealize the condition of the information communicating between controllers and plants in \cite{seron1997fundamental,chen2000limitations,freudenberg2003fundamental,middleton2004tracking,11Wang}. Some additional essential issues still remain unsolved, e.g., best design and accomplishable tracking capability of NCSs with communication restrictions, (e.g. quantization, bandwidth, time-delays, data packet and channel noise).
\par
Whether the performance of the system can be achieved is a crucial problem in control system design. The performance limitation, as the lower bound of the performance that the system can achieve, has important guiding significance for the system design \cite{chen2003guest}. For a NCSs, the optimal performance is affected by the control system and network constraints \cite{fang2017design}. How to obtain the quantitative relationship between the optimal performance and the internal characteristics of the control system or the various network constraints, becomes the key to the performance limitation of NCSs. With the development of NCSs, the study of its performance limitations has attracted extensive interest from researchers. In the last ten years, the scientists have been expanding the studies in the tracking performance limitation to NCSs. For example, in terms of signal to noise ratio (SNR) capability limitation, the SNR restricted communication paths are considered in \cite{07Braslavsky,rojas2008fundamental}, which demonstrates the restrictions of the quality of stabilizing an unstable plant. The restriction is considered to be over a SNR limited path by finite-dimensional LTI feedback. Then, Rojas et al. \cite{09Rojas} extended the results, he studies the SNR fundamental limitation as considering the control channels of a LTI feedback loop with an additive coloured Gaussian noise (ACGN) channel. In the optimal performance of NCSs, the tracking capability of continuous-time, MIMO, LTI systems with a channel noise in output feedback are put into consideration \cite{10Ding}. The results demonstrate the possible way the AWGN worsens the tracking performance. Goodwin et al. \cite{10Goodwin} presents a summary listing recent achievements in NCSs when underlying the additive noise model methodology, also point to several open problems in this area. Guan et al. \cite{13Guan} works on the case of the tracking performance limitations of MIMO NCSs with ACGN channel in downlink channel. The tracking performance limitations are investigated in \cite{zhan2013optimal} and \cite{zhan2015optimal} for NCSs in the feedback path with packet dropouts or time delay, which theses papers are concerned with performance limitations are determined by the internal structure of the plant and communication parameters, and a novel design method of controller is obtained by the frequency-domain analysis. Then, the case of performance limitations of MIMO LTI systems with communication noises and packet dropouts has been promoted \cite{13GuanJiang}. In \cite{2014Latorre}, the impact of adding additional energy constrained control inputs to the plant is studied on the accomplishable closed loop performance. Li et al. \cite{Chen16} investigated the system's stabilizability as well as tracking performance under the additional white noise (AWN) channel power constraint. Qi et al. \cite{qi2017control} give a fundamental conditions of stabilizability. %Fang, etc \cite{fang2017design} provide the fundamental limits of disturbance attenuation achievable by networked feedback from an information theory perspective.
In recent years, we have also obtained some optimal performance results of NCSs, such as upstream and downstream channels with constraints \cite{chen2016optimal}, channel energy constaint \cite{wang2016trade}, some novel trade-off factors and constraint channels \cite{chen2017fundamentala}, discrete-time(DT) systems with quantization \cite{chen2017fundamentalb}, AWGN fading channels \cite{chen2018performance}, SIMO systems with packet-dropouts\cite{2018Jiang}. In spite of the significant progress on optimal performance studies, there are still many gaps in the optimal performance of NCSs.
\par
Tracking performance limitations issues have been studying about ten years in regard to finite-dimensional, LTI NCSs. Some existing literatures above tracking performance limitations of NCSs concentrate upon simulated path models emphasizing specific fields of the wholesome case, such as \cite{10Goodwin,13Guan,14ZhanGuan,Chen16,chen2017fundamentalb,2018Jiang} etc. However, multiple communication constraints are often encountered in general practice, and the performance limitations problem may become more complex and realistic. We focus on understanding the inherent relationship between multiple communication constraints and system performance limitations for MIMO NCSs, and explore whether communication constraints have a coupling relationship when jointly affecting performance. Moreover, we further analyze the difference affects of performance limitations between SISO NCSs and MIMO NCSs, and the influence when each channel has different power distribution for MIMO NCSs. For the networked communication constraints, we focus on some basic network constraints in NCSs, including communication noises, bandwidth, encode-decode and quantification, which are generally considered to be always present when the signals are communicated in the NCSs. They are different from the network-induced communication constraints, such as time delay, packet loss, and so on. Almost any input signals are inevitably subject to random interference when these signals are input a control system or when a control system tracks a given reference signal in practice. Therefore, interference noises of the reference signal (or call reference noises) is considered as the input signal in our control system performance model. This is also widely used method when studying control system performance in recent years, such as \cite{10Ding,12GuanZhan,13Guan,14ZhanGuan,Chen16,chen2017fundamentalb,2018Jiang} et al. In addition, we adopt the 2DOF controller, which can be recognized as a controller structure that achieves better performance \cite{chen2000limitations,11Wang}, because this kind of controller structure has more design freedom.
\par
In this paper, we investigates the tracking performance limitations for MIMO NCSs with multiple communication constraints, and the plant has unstable poles and NMP zeros. The contributions and significance lie in the following several folds: \par
(i) Multiple basic communication constraints are considered simultaneously (including communication noises, bandwidth, encode-decode and quantification). They are inherent factors in network communication different from network-induced factors (such as delay and packet loss).\par
(ii) The relationship among tracking performance limitations, internal characteristics of the plant, reference noises and communication constraints are presented in quantitatively.
Tracking performance limitations depend on characteristics of plant (unstable poles, NMP zeros and their directions), and network constraints (the reference noise power distribution, the communication noise power distribution, the quantization noise power distribution, and their power distribution directions, and the transform encoder-decoder allocation and transform bandwidth allocation, and their allocation directions, and the NMP zeros of bandwidth and the MP part of bandwidth).  \par
(iii) Different from the classic performance limitation results, the results indicate the performance limitations of MIMO NCSs are also related to the angles between NMP zeros of plant and reference noise power allocation, the angles between unstable poles of plant and communication constraint allocation direction, which can greatly affect the performance limitations. Furthermore, the performance limitations are given under some different communication constraint typical combinations. In addition, the NMP bandwidth model in this paper is considered as a more universal model.
\par
This paper is organized as follows. Section \ref{se2} provides a overall subsequent development with a preliminary background. Section \ref{se3} defines the network control model and formulates the case of the tracking performance. Section \ref{se4} first studies the best accomplishable tracking performance for SISO NCSs, and then investigated the tracking performance limitations for MIMO NCSs,  while Section \ref{se6} carried out some numerical simulations. This paper is concluded in Section \ref{se7}.

\section{PRELIMINARIES}\label{se2}

The notations used are demonstrated as follows. ${\bar z}$ denotes complex conjugate of any complex number $z$. $\bm u^T$ denotes the transpose of a vector $\bm u$, and conjugate transpose of $\bm u$ is denoted by $\bm u^H$. Let $\mathcal{E}(\cdot)$, $\mathop{\rm Tr}(\cdot)$ and $\mathop{\rm Re}(\cdot)$ denote expectation operator, trace operator and real part operator, correspondingly. $A$ is used to express any matrix, $A^T$, $A^H$ and $A^\dag$ denote the transpose, the conjugate transpose and the right-inverse of matrix $A$, respectively. $\emph{diag}\{z_i\}$ denotes a compatible dimension diagonal matrix, $z_i$ represents the corresponding element of the $i$-th row and the $i$-th column. We suppose all of the vectors, together with matrices have compatible dimensions. In addition, $C_ -  : = \left\{ {s:{\mathop{\rm Re}} (s) < 0} \right\}$, indicates the open left halves of the complex plane, while $C_ +  : = \left\{ {s:{\mathop{\rm Re}} (s) > 0} \right\}$ indicates the right correspondingly. And we refer
$|\cdot|$ to the absolute value or modulus, $\|\cdot\|$ to the Euclidean norm, $\|\cdot\|_F$ to the Frobenius norm. Moreover, the class of all stable and the matrices of proper rational transfer function are written as ${\rm \bf R\mathcal{H}}_\infty$. In regard to the nonzero vectors $\bm \alpha$ and $\bm \beta$, we give a description as $\cos\angle(\bm \alpha, \bm \beta)={|\bm \alpha^H \bm \beta|}/{(\|\bm \alpha\|\|\bm \beta\|)},$ where $\angle(\bm \alpha, \bm \beta)$ represents the main angle of the two subspaces which are spanned by $\bm \alpha$ and $\bm \beta$.
The Hilbert Space is written as
\begin{align*}
\mathcal{L}_2 : =\{ P:P(s)~&{\rm{measurable}}~{\rm{in}}~~{\rm{ }}C_0,\\
&\| P \|_{\rm{2}}^{\rm{2}} {\rm{: = }}\frac{{\rm{1}}}{{{\rm{2}}\pi }}{\rm{ }}\int_{{\rm{ - }}\infty }^\infty  \| {P(j\omega)} \|_F^2d\omega
 < \infty  \},
\end{align*}
when we define the inner product as follows:
\begin{align*}
\langle {P_1, P_2} \rangle : = \frac{1}{2\pi }\int_{-\infty }^\infty {\rm Tr}\{P_1^H (j\omega)P_2(j\omega)\}d\omega.
\end{align*}
Denote $\mathcal{L}_2$ which admits an orthogonal decomposition (OD) into two subspaces
\begin{align*}
\mathcal{H}_2 : =\Big\{ P:&P(s)~{\rm{ }}{\emph{analytic}}~{\rm{ }}in~{\rm{ }}C_ +  {\rm{,}}\\
&\|P\|_{\rm{2}}^{\rm{2}} {\rm{: = }}\mathop {{\rm{sup}}}\limits_{{\sigma} {\rm{ > 0}}} \frac{{\rm{1}}}{{{\rm{2}}\pi }}{\rm{ }}\int_{{\rm{ - }}\infty }^\infty \| {P({\sigma}  + j\omega)} \|_F^2 d\omega < \infty  \Big\},
\end{align*}
and
\begin{align*}
\mathcal{H}_2 ^ \bot  : = \Big\{ P:&P(s)~{\rm{ }}\emph{analytic}~{\rm{ }}in~{\rm{ }}C_-{\rm{,}}\\
&\| P \|_{\rm{2}}^{\rm{2}} {\rm{: = }}
\mathop {{\rm{sup}}}\limits_{{\sigma} {\rm{ < 0}}} \frac{{\rm{1}}}{{{\rm{2}}\pi }}{\rm{ }}\int_{{\rm{ - }}\infty }^\infty  \| {P({\sigma}  + j\omega)} \|_F^2 d\omega < \infty  \Big\}.
\end{align*}
Obviously, when it comes to any $P_1 \in \mathcal{H}_2$ as well as $P_2 \in \mathcal{H}_2^ \bot$, $<P_1,P_2 > = 0$.

%$$fdf\eqno{(8)}$$

\section{PROBLEM FORMULATION}\label{se3}

\begin{figure}[ht]
 \centering
  \includegraphics[width=8cm]{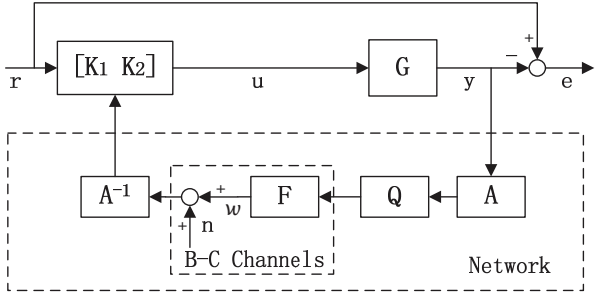}
  \captionsetup{justification=centering}
  \caption{\small{The MIMO Plant Control Scheme.}}\label{fig1a}
  \end{figure}

The feedback structure is provided by Fig. \ref{fig1a}, in which $G$ acts as the plant. General bandwidth communication channels (B-C Channels)\cite{rojas2008fundamental} are considered. This paper models the bandwidth restriction by the low pass transfer function $F=diag\{f_i\},(i=1,2,\cdots,m)$ (look up to, e.g., \cite{09Rojas} as well as the notes wherein). $[K_1~~K_2]$ represents a 2DOF, while $\mathcal{Q}$ represents uniform quantizer, $A$ represents the term encoder, and $A^{-1}$ represents the decoder, respectively (e.g., \cite{10Goodwin,Chen16} and the references wherein). Let $A=diag\{\lambda_1,\lambda_2,\cdots,\lambda_m\}$. It is suggested that signal $\bm r$ indicates the reference noises, signal $\bm q$ indicates the quantization noises, and signal $\bm n$ indicates the channel noises (or called communication noises), accordingly. Assume that the quantizer does not overload, and the quantized values of the outputs will be treated as an input affected by an additive noise \cite{08Widrow}. The quantization noise $\bm q=[q_1, q_2,\cdots, q_m]^T$ in dissimilar channels are statistically self-reliant and are not related to one another, and we consider $q_i,(1\leq{i}\leq{m})$ as a procedure of an AWN which distributed uniformly over $[{-\Delta_i}/{2},~{-\Delta_i}/{2}]$. $\Delta_i$ is the quantization interval as is known to all, and $\Delta_i=2M_i/(2^{b_i}-1)$, where $b_i$ is the amount of bits which is distributed for channel transmission. $[-M_i, M_i]$ is the overall quantiser range, while considering $M_i\in{R}$ providing that the chance of overflow is small.
%and the value of $b_i$ is assumed such that the number of quantisation levels is high.
The variance of $q_i$ can be defined $\sigma_{qi}^2=\Delta_i^2/12$ as reference \cite{08Gallager}.
The path model is generated by the bandwidth-limited (BL) AWN path. We've noticed that a uniformly distributed AWN is a well-recognized system of methods which is widely used in signal processing literatures \cite{zamir1992universal}.
It is considered that the channel communication noises $\bm n=[n_1, n_2,\cdots, n_m]^T$ and the reference noises $\bm r=[r_1, r_2,\cdots, r_m]^T$
are zero mean i.i.d. additional white noises (GWN) which obtains power spectral density (PSD) of ${\sigma_r}_i^2$ and ${\sigma_n}_i^2$.
\begin{remark}
We assume the reference noises are GWN, because we believe the reference noises are always exist at the signal input port, such as the reference \cite{Chen16} also adopt similar signals. If it is assumed that this signals only occur or exist when the reference signals are input, it can be set as step random reference noises \cite{13Guan}, and the corresponding conclusions can be deduced in parallel.
\end{remark}
\par
In addition, these signals are transmitted in different channels so that they are not related to one another when being analyzed from statistical data. And when it comes to a scalar channel, this paper gives $\mathcal{E}\{|r|^2\}=\sigma_r^2$, $\mathcal{E}\{|n|^2\}=\sigma_n^2$ and  $\mathcal{E}\{|q|^2\}=\sigma_q^2$. Assuming the signals $\bm r, \bm n$
and $\bm q$ are uncorrelated. Let
\begin{align*}
\mathcal{U}&=diag\{{\sigma_r}_i\},(i=1,2,\cdots,m),~~\Phi_r=\|\mathcal{U}\|_F^2,\\
\mathcal{V}&=diag\{{\sigma_n}_i\},(i=1,2,\cdots,m),~~\Phi_n=\|\mathcal{V}\|_F^2,\\
\mathcal{Q}&=diag\{{\sigma_q}_i\},(i=1,2,\cdots,m),~~\Phi_q=\|\mathcal{Q}\|_F^2,
\end{align*}
and it is instructive to represent the reference noise \emph{power distribution direction} (PD-Direction), communication noise PD-Direction and quantization noise PD-Direction by unitary vectors
\begin{align*}
\bm \upsilon_r&=[\sigma_{r_1}^2, \sigma_{r_2}^2,\cdots,\sigma_{n_m}^2]^T/{\Psi_r},\\
\bm \upsilon_n&=[\sigma_{n_1}^2, \sigma_{n_2}^2,\cdots, \sigma_{n_m}^2]^T/{\Psi_n},
\end{align*}
\begin{align*}
\bm \upsilon_q&=[\sigma_{q_1}^2, \sigma_{q_2}^2,\cdots, \sigma_{q_m}^2]^T/{\Psi_q},
\end{align*}
where
$
\Psi_r=\|\mathcal{U}^H\mathcal{U}\|_F,
\Psi_n=\|\mathcal{V}^H\mathcal{V}\|_F,
\Psi_q=\|\mathcal{Q}^H\mathcal{Q}\|_F.
$
The $\Psi_r$, $\Psi_n$ and $\Psi_q$ are called the reference noise \emph{power distribution square-sum block} (PD-SSB), the communication noise PD-SSB and the quantization noise PD-SSB.
In addition, let
\begin{align*}
\Psi_{F}&=\left\|[|f_1|^{-2}, |f_2|^{-2},\cdots,|f_m|^{-2}]^T\right\|_F,~~\\
\Psi_{A}&=\left\|[\lambda_1^{-2}, \lambda_2^{-2},\cdots,\lambda_m^{-2}]^T\right\|_F,~~\\
\bm \upsilon_{F}&=[|f_1|^{-2}, |f_2|^{-2},\cdots,|f_m|^{-2}]^T/\Psi_{F},~~\\
\bm \upsilon_{A}&=[\lambda_1^{-2}, \lambda_2^{-2},\cdots,\lambda_m^{-2}]^T/\Psi_{A},
\end{align*}
$\Psi_{F}$ and $\Psi_{A}$ are called bandwidth \emph{transform allocation square-sum block} (TA-SSB) and encoder-decoder TA-SSB. $\bm \upsilon_{F}$ and $\bm \upsilon_{A}$ are called bandwidth \emph{transform allocation direction} (TA-direction) and encoder-decoder TA-direction, they are the unitary vectors.
%\begin{center}
%\includegraphics[width=8.5cm]{MIMO.eps}
%\includegraphics[height=3.1cm,width=7.6cm]{Fig.1.pdf}
%{\footnotesize \centerline{{\bf Fig. 1.} The MIMO Plant Control Scheme}}
%\end{center}

In this paper, we use the integrated squared error (ISE) criterion to analyze the tracking performance:
\begin{align}
J = :\mathcal{E}\left\{ {\left\|\bm e(s)\right\|_2^2 } \right\}=\mathcal{E}\left\{ {\left\|\bm y(s)-\bm r(s) \right\|_2^2 } \right\}.\label{1}
\end{align}
where $\bm e(s)$ is the tracking error.\par
Our relevant work is ascertaining the accomplishable performance limitations through stabilizing the compensators in $\mathcal{K}$, which is indicated by
$J^ * : = \mathop {\inf }_{\mathop {K}\in\mathcal{K} } J.$
The related controlled plant includes NMP zeros as well as unstable poles.
\par
In this paper we resort to factorizations, and consider that the coprime factorization of $G$ and $FG$ can be provided by
\begin{align}
G&=NM^{ - 1}={{\tilde M}^{ - 1}}\tilde N,\label{eq1a}\\
FG&= N_FM_F^{ - 1}={{\tilde M}_F^{ - 1}}\tilde N_F,\label{eq2a}
\end{align}
where $N, \tilde N, N_F, \tilde N_F, M, {\tilde M}, M_F, {\tilde M}_F\in RH_\infty$, $N_F=FN$, $M_F=M$.
Because $F$ is stable, ${\tilde M}_F$ and ${\tilde M}$ have the same NMP zeros, but the NMP zeros direction has changed due to the impact of the bandwidth model.
$N_F, \tilde N_F, M_F, {\tilde M}_F$ are satisfied with the double Bezout identity\cite{87Francis}:
\begin{equation}\label{eq1}
    \left[ {\begin{array}{cc}
{\tilde X }&{ - \tilde Y}\\
{ - \tilde N_F}&{\tilde M}_F
\end{array}} \right]\left[ {\begin{array}{*{20}{c}}
M_F&Y\\
N_F&X
\end{array}} \right] = I,
\end{equation}
for $X, Y, \tilde X, \tilde Y  \in {\rm \bf R\mathcal{H}}_\infty$.
As is known to all, by the the Youla parameterization \cite{92Doyle,chen2016optimal},
%it is possible to denote every stabilizing one-degree-of-freedom (1DOF) compensator(this is a commonly used controller structure) $K$ as
%\begin{align}
%{\mathcal{K}_{OF}} :
%= \Big\{ &K:K =  -(Y-M_FQ)(X-N_FQ)^{-1}\nonumber\\
%-&(\tilde X - Q\tilde{N}_F)^{ - 1}(\tilde{Y}-Q\tilde{M}_F),
%Q, \in {\rm \bf R\mathcal{H}}_\infty\Big\},
%\end{align}
it is also possible to denote every stabilizing \emph{two-degree-of-freedom} (2DOF, or called Two-parameter) compensator $K$.
The set of all stabilizing 2DOF is characterized by
\begin{align}
{\mathcal{K}} : = \Big\{ K:K &= [K_1 {\rm{  }}~~ K_2 ]= (\tilde X - R\tilde N_F)^{ - 1}\nonumber\\
&\left[ {\begin{array}{*{20}c}
   Q & {\tilde Y - R\tilde M}_F  \\
\end{array}} \right], Q, R \in {\rm \bf R\mathcal{H}}_\infty\Big\},\label{eq4a}
\end{align}
where $Q$ and $R$ are the controller's parameters, which can be free design.
\par
The unstable poles as well as the NMP zeros of $G(s)$ are written as $p_k$ ($k = 1, \ldots ,n_p$, and $z_k$ ($k = 1, \ldots ,n_z$, respectively. We use $s_k$, ($k = 1, \ldots ,n_f$ to denote NMP zeros of $F(s)$. And $p_k$ $s_k$ and $z_k$ are \emph{simple}.  Owning to $F(s)$ is a diagonal matrix, it is well known that $F(s)$ can be factorized as
\begin{align}
F(s) =L_f(s)F_m (s)=F_m (s)L_f(s),\label{eq6}
\end{align}
where $F_m (s)$ stands for the MP part of $F(s)$, and $F_m (s)$ are diagonal, denote
%\begin{align}
%L_f (s)&=\emph{diag}[l_{f1},l_{f2},\cdots,l_{fm}]^T,\label{eqn1}\\
$F_m (s)=\emph{diag}[f_1^{(m)},f_{2}^{(m)},\cdots,f_m^{(m)}]^T.$
%\end{align}
$L_f(s)$ is all pass factor, and it is reliable for us to write it as
$
L_f(s):=\prod_{i= 1}^{n_f}{{L_f}_i}(s),
$ where
\begin{align}\label{eq7aa}
   {{L_f}_i}(s):=&{[{\bm \eta}_{fi}~~{U}_{fi}]
\left[
  \begin{array}{cc}
    \frac{s-s_i}{s+\bar{s}_{i}} & 0 \\
    0 & I
  \end{array}
\right]
  \left[
    \begin{array}{c}
      {\bm \eta}_{fi}^H \\
      {U}_{fi}^H
    \end{array}
  \right]},
\end{align}
where ${U}_{fi}$ is a matrix whose columns, accompanied by ${\bm \eta}_{fi}$, creates an orthonormal basis of the corresponding Euclidean space, i.e.,
${\bm \eta}_{fi}{\bm \eta}_{fi}^H+{U}_{fi}{U}_{fi}^H=I$.
\par
\begin{remark}
The NMP bandwidth model in this paper is a more universal model in performance analysis of MIMO NCSs. Some literatures have studied some special cases. For example, the stable and minimum bandwidth model is considered in \cite{14ZhanGuan} in SISO NCSs, the SNR performance limitations of SISO system are analyzed with NMP bandwidth model in \cite{rojas2008fundamental,09Rojas}, the NMP bandwidth model is studied in \cite{13Guan} with same constraints in each channels of MIMO NCSs.
\end{remark}
\par
Furthermore, we factorize coprime factors $N(s)$, $\tilde{M}(s)$ and $\tilde{M}_F(s)$ as
\begin{align}
N &= LN_{m},~~~~~\tilde{M} = \tilde{M}_{m}\tilde{B},\label{eq6a}\\
N_F &= L_FN_{Fm},~\tilde{M}_F = \tilde{M}_{Fm}\tilde{B}_F\label{eq7}
\end{align}
in which $N_{m}$ and $N_{Fm}$ represents the MP part of $N$ and $FN$, and $\tilde{M}_{m}$ represents the MP part of $\tilde{M}$, $\tilde{M}_{Fm}$ represents the MP part of $\tilde{M}_F(s)$ respectively. $L, L_F, \tilde{B}$ and $\tilde{B}_F$ are all pass factors. And it is possible to write for $L, \tilde{B}$ and $\tilde{B}_F$ as follows.
$
L:=\prod_{i=1}^{n_z}{L_i},
 \tilde{B}:=\prod_{i=1}^{n_p}\tilde{B}_{n_p-(i-1)},~~
 \tilde{B}_F:=\prod_{i=1}^{n_p}\tilde{B}_{Fn_p-(i-1)},
$
 and
\begin{align}
   {L_i}(s):=&{[{\bm \eta}_i~~{U}_i]
\left[
\begin{array}{cc}
   \frac{s-z_i}{s+\bar{z}_i}& 0 \\
    0 & I
  \end{array}
\right]
  \left[
    \begin{array}{c}
     {\bm \eta}_i^H \\
     {U}_i^H
    \end{array}
  \right]},\label{eq4}\\
 {\tilde{B}_{i}}(s):=&{[{\bm \omega}_i~~{W}_i]
\left[
  \begin{array}{cc}
    \frac{s-p_i}{s+\bar{p}_i} & 0 \\
    0 & I
  \end{array}
\right]
  \left[
    \begin{array}{c}
      {\bm \omega}_i^H \\
      {W}_i^H
    \end{array}
  \right]},\label{eq5}
\\
 {\tilde{B}_{Fi}}(s):=&{[{\hat{\bm \omega}}_i~~{\hat{W}}_i]
\left[
  \begin{array}{cc}
   \frac{s-p_i}{s+\bar{p}_i} & 0 \\
    0 & I
  \end{array}
\right]
  \left[
    \begin{array}{c}
      {\hat{\bm \omega}}_i^H \\
      {\hat{W}}_i^H
    \end{array}
  \right]},\label{eq5a}
\end{align}
where ${\bm \eta}_i{\bm \eta}_i^H+{U}_i{U}_i^H=I$, ${\bm \omega}_i{\bm \omega}_i^H+{W}_i{W}_i^H=I$ and ${\hat{\bm \omega}}_i{\hat{\bm \omega}}_i^H+{\hat{W}}_i{\hat{W}}_i^H=I$, respectively.  $\bm \eta_i$, $\bm \omega_i$ and $\hat{\bm \omega}_i$ are called zero direction, pole direction and pole \emph{bandwidth interference direction} (BI-Direction), respectively. And, $\eta_i$, $\bm \omega_i$ and $\hat{\bm \omega}_i$ are unitary vectors. Let
\begin{align*}
\Psi_{\check{\eta}_i}=&\left\|[\eta_{i1}^2,\eta_{i2}^2,\dots,\eta_{i1}^2]^T\right\|_F,
\bm {\check{\eta}}_i=[\eta_{i1}^2,\eta_{i2}^2,\dots,\eta_{i1}^2]^T/\Psi_{\check{\eta}},\\
\Psi_{\check{\omega}_i}=&\left\|[\omega_{i1}^2,\omega_{i2}^2,\dots,\omega_{i1}^2]^T\right\|_F,
\bm {\check{\omega}}_i=[\omega_{i1}^2,\omega_{i2}^2,\dots,\omega_{i1}^2]^T/\Psi_{\check{\omega}},\\
\Psi_{\check{\hat{\omega}}_i}=&\left\|[\hat{\omega}_{i1}^2,\hat{\omega}_{i2}^2,\dots,\hat{\omega}_{i1}^2]^T\right\|_F,
\bm {\check{\hat{\omega}}}_i=[\hat{\omega}_{i1}^2,\hat{\omega}_{i2}^2,\dots,\hat{\omega}_{i1}^2]^T/\Psi_{\check{\hat{\omega}}},
\end{align*}
where $\Psi_{\check{\eta}_i}, \Psi_{\check{\omega}_i}$ and ${\Psi}_{\check{\hat{\omega}}_i}$ are called zero \emph{direction square-sum block} (DSSB), pole DSSB and pole \emph{bandwidth interference direction square-sum block} (BI-DSSB), respectively. $\bm {\check{\eta}}_i, \bm {\check{\omega}}_i$ and $\bm {\check{\hat{\omega}}}_i$ are called zero \emph{transform direction} (T-Direction), pole T-Direction and pole \emph{bandwidth interference transform direction} (BI-T-Direction), respectively. Obviously, $\bm {\check{\eta}}_i, \bm {\check{\omega}}_i$ and $\bm {\check{\hat{\omega}}}_i$ are unitary vectors.
\vspace{5pt}
\begin{lemma}\label{le1}
Make $L$ and $B$ be determined similarly by (\ref{eq4}) and (\ref{eq5}), and $z_i$ and $p_i$ with multiplicity $1$, for any $X, Y \in {\rm \bf R\mathcal{H}}_\infty$, the equations
\begin{align*}
L^{ - 1}Y=&T +\sum\limits_{i = 1}^{n_z }{L}_{Oi}(z_i)(L_i^j) ^{ - 1} {L}_{Ii}(z_i)Y(z_i),\\
X\tilde{B}^{ - 1}=&S +\sum\limits_{i = 1}^{n_p }X(p_i)\tilde{B}_{Ii}(p_i)(\tilde{B}_i^j) ^{ - 1} \tilde{B}_{Oi}(p_i)
\end{align*}
hold for some $S, T \in {\rm \bf R\mathcal{H}}_\infty$, where
\begin{align*}
{L}_{Ii}(s)&=\prod_{k=1}^{i-1}L_{k}(s),~{L}_{Oi}(s)=\prod_{k=i+1}^{n_z}L_{n_p-k}(s),\\
\tilde{B}_{Ii}(s)&=\prod_{k=1}^{i-1}\tilde{B}_{i-k}(s), ~
\tilde{B}_{Oi}(s)=\prod_{k=i+1}^{n_p}\tilde{B}_{n_p+(i+1-k)}(s).
\end{align*}
\end{lemma}
\begin{IEEEproof}
Similar to the proof of lemme 3 in our work \cite{13Guan} and combined partial fraction expansion technique, it is easy to prove.
\end{IEEEproof}
%The lemma can be easily proved by mathematics inductive method.
\vspace{5pt}
The prerequisite work of our research is assuming that the plant transfer function matrix is right invertible. The premise is indispensable for the purpose of achieving asymptotic tracking\cite{11Wang,Chen16}. We look into 2DOF and calculate the tracking error energy by an ISE criterion. The minimal tracking error caused by the reference noise, additive communication noise, quantization noise, bandwidth and encoder-decoder in the feedback channel are given in the following sections.

\section{TRACKING PERFORMANCE LIMITATIONS of NCSs}\label{se4}
This paper firstly investigates the tracking performance limitations of SISO NCSs with multiple communication constraints. The controlled plant is unstable and NMP system. And then, we further investigated the performance limitations of MIMO NCSs with multiple communication constraints. And the influences of various communication constraints on performance limitations is analyzed and discussed. In this paper, when it comes to a scalar channel, we donate $f=f_1, \lambda=\lambda_1, f^{(m)}=f_1^{(m)}$.
\par

\subsection{Tracking Performance Limitations of SISO NCSs with Multiple Communication Constraints}\label{subse1}
\begin{figure}[ht]
 \centering
  \includegraphics[width=8cm]{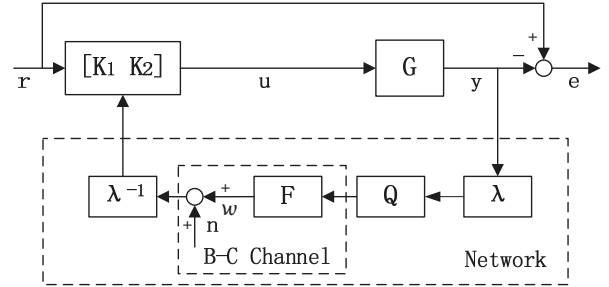}
  \captionsetup{justification=centering}
  \caption{\small{The SISO Plant Control Scheme.}}\label{fig.2}
  \end{figure}

In this section, in order to clearly examine the influences between tracking performance limitation and system characteristics, network constraints, the reference noise on system performance, the SISO NCS is considered. Considering the Fig. \ref{fig.2}, we can get the corollary as follows.
%\begin{figure}[H]
%  \centering
%  \includegraphics[width=11cm]{SISO.eps}
%  \caption{The SISO Plant Control Scheme}
%  \label{fig1b}
%\end{figure}
\vspace{5pt}
\begin{theorem}\label{thm1}
Considering the SISO system, the NCS structural model is demonstrated in Fig. \ref{fig.2}.
The reference signal $\bm r$, the communication noise $\bm n$ and the quantization noise $\bm q$ are ought to be a zero mean i.i.d. GWN with PSD ${\sigma_r}^2, {\sigma_q}^2$ and ${\sigma_n}^2$. The encoder $A=\lambda$. Assume that $p_k~(k = 1, \ldots ,n_p )$ are unstable poles of $G(s)$, and $z_k~(k = 1, \ldots ,n_z )$ and $s_k~(k = 1, \ldots ,n_f )$ are NMP zeros of $G(s)$ and $F(s)$, respectively. Then,
\begin{align*}
J^*=2{\sigma_r}^2\sum\limits_{i = 1}^{{n_z}} {\mathop{\rm Re}} ({z_i})+\sum\limits_{i,j = 1}^{n_p}{W}_{sys}^{i,j}{W}_{net}^{i,j},
\end{align*}
where
\begin{align*}
W_{sys}^{i,j}=&\frac{{4{\mathop{\rm Re}} ({p_i}){\mathop{\rm Re}} ({p_j})}}{{{{\bar p}_i} + {p_j}}}
\prod_{k=1}^{n_z}\frac{\bar p_i+ z_k}{\bar p_i-\bar z_k}
\prod_{k=1,k\neq{}i}^{n_p}\frac{\bar p_i+p_k}{\bar p_i-\bar p_k}\\
&\times\prod_{k=1}^{n_z}\frac{p_j+\bar z_k}{p_j-z_k}
\prod_{k=1,k\neq{}j}^{n_p}\frac{p_j+\bar p_k}{p_j-p_k}\\
W_{net}^{i,j}=&\lambda^{-1}\Big(\big(\sigma_n/\bar{f}^{m}(p_i)+\sigma_q\big)(\sigma_n/f^{m}(p_j)+\sigma_q)\Big)\\
&\times\prod_{k=1}^{n_f}\frac{\bar p_i+ s_k}{\bar p_i-\bar s_k}
\prod_{k=1}^{n_f}\frac{p_j+\bar s_k}{p_j-s_k}.
\end{align*}
\end{theorem}
\begin{IEEEproof}
 The transfer function of $\bm n$, $\bm q$ and $\bm r$ to $\bm y$ is constructed as follows
$
\bm y = G{K_1}\bm r + G{K_2}{\lambda^{ - 1}}[\bm n + F(\bm q + \lambda{\bm y}),
$
then
$
\bm y={(1 - G{K_2}F)^{ - 1}}G\big({K_1}\bm r + {K_2}{ \lambda^{ - 1}}(\bm n+F\bm q)\big).
$
\par
On the basis of the coprime factorization (\ref{eq1a})(\ref{eq2a}) and the Youla parameterization (\ref{eq4a}), it is possible to rewrite the above-mentioned transfer function as follows:
\begin{align*}
\bm y=& G{(1 -{K_2}FG)^{ - 1}}\big({K_1}\bm r + {K_2}{ \lambda^{ - 1}}(\bm n+F\bm q)\big)\\
=& G\Big(1 -(\tilde{X}-R\tilde{N}_F)^{-1}(\tilde{Y}-R\tilde{M}_F)N_FM_F^{-1}\Big)^{-1}\\ &\times\Big((\tilde{X}-R\tilde{N}_F)^{-1}Q\bm r + (\tilde{X}-R\tilde{N}_F)^{-1}\\
&\times(\tilde{Y}-R\tilde{M}_F){ \lambda^{ - 1}}(\bm n+F\bm q)\Big)
\end{align*}
\begin{align*}
=& GM_F\Big( (\tilde{X}-R\tilde{N}_F)M_F -(\tilde{Y}-R\tilde{M}_F)N_F\Big)^{-1}\\
&\times\Big(Q\bm r + (\tilde{Y}-R\tilde{M}_F){ \lambda^{ - 1}}(\bm n+F\bm q)\Big).
\end{align*}
By using the double Bezout identity (\ref{eq2a}), noting $M_F=M$ we have
$
\bm y= NQ\bm r + N(\tilde Y - R\tilde M_F){ \lambda^{ - 1}}(\bm n+F\bm q).
$
On the basis of the performance index (\ref{1}) and $\bm r, \bm n, \bm q$ are uncorrelated, and it can be designed $\tilde{M}_F=\tilde{M}$ for SISO system, it follows that
\begin{align*}
J=&\mathcal{E}\left\{ {\left\| \bm y(s)-\bm r(s) \right\|_2^2 } \right\}\\
=&\left\| {(1 - NQ){\sigma_r}} \right\|_2^2 + \left\| N_m(\tilde{Y}-R\tilde{M})\lambda^{ - 1}(\bm n+F\bm q)\right\|_2^2.
\end{align*}
we define
\begin{align}
J_1=&\left\| {(1 - NQ){\sigma_r}} \right\|_2^2,\label{eq1aa}\\
J_2=&\left\| N_m(\tilde{Y}-R\tilde{M})\lambda^{ - 1}(\bm n+F\bm q)\right\|_2^2.\label{eq2aa}
\end{align}
\par
Because the controller parameters $Q$ and $R$ are independently designable parameters, we have
\[J^*=\inf_{K\in\mathcal{K}}J=\inf_{Q\in{{\rm \bf R\mathcal{H}}_\infty}}J_1+\inf_{R\in{{\rm \bf R\mathcal{H}}_\infty}}J_2.\]
\par
First of all, for $J_1$, noting equations (\ref{eq1aa}), it holds that
\begin{align*}
J_1^*=&\inf_{Q\in {\rm \bf R\mathcal{H}}_\infty}\left\| {(1 - NQ){\sigma_r}} \right\|_2^2\\
=&\inf_{Q\in {\rm \bf R\mathcal{H}}_\infty}
\left\| (L^{-1}-1)+(1 - N_{m}Q))\sigma_r \right\|_2^2\nonumber
\end{align*}
Owning to $L^{-1}(0)=1$ and $(L^{-1}-1)\sigma_r\in{\mathcal{H}_2^\perp}$, and $Q$ can be select so that
$(I - N(0)Q(0))\sigma_r=0$, we have $(1 - N_{m}Q)\sigma_r\in{\mathcal{H}_2}$, then
\begin{align*}
J_1^*=&\left\| (L^{-1}-1)\sigma_r\right\|_2^2+\inf_{Q\in {\rm \bf R\mathcal{H}}_\infty}\left\|(1 - N_{m}Q)\sigma_r \right\|_2^2\nonumber
\end{align*}
Because $N_m$ is MP part the coprime factor $N$, we can design $Q=N_m^\dag\sigma_r^{-1}$, then
$\inf_{Q\in {\rm \bf R\mathcal{H}}_\infty}\left\|(I - N_{m}Q)\sigma_r \right\|_2^2=0.$
Thus,
\begin{align}
J_1^*=&\left\| (L^{-1}-I)\sigma_r\right\|_2^2\nonumber\\
=&\sum\limits_{i = 1}^{{n_z}}\left\| \frac{2{\rm Re}(z_i)}{s+\bar{z_i}}\sigma_r\right\|_2^2\nonumber\\
=&2{\sigma_r}^2\sum\limits_{i = 1}^{{n_z}} {\mathop{\rm Re}} ({z_i}).\label{eq8}
\end{align}
\par
%Secondly, for $J_2$, noting equation (\ref{eq6})-(\ref{eq7}), similar shown in \cite{Chen16} that there exists some $Z\in{\mathbf{R}}\mathcal{H}_\infty$, such that $N_F\hat{Y}\tilde{B}_F^{-1}=\tilde{B}^{-1}_F+Z$,
%then $N\hat{Y}=f^{-1}+f^{-1}Z\tilde{B}$. From the equation (\ref{eq2aa}), we have
%\begin{align*}
%{J_2} =& \left\| (f^{-1}+f^{-1}Z\tilde{B}-NR\tilde{M})H \right\|_2^2\\
%=&\left\| (f^{-1}B^{-1}-I\|+\|f^{-1}Z-NR\tilde{M}_mH \right\|_2^2\\
%=&\left\| (f^{-1}HB^{-1}+f^{-1}Z-NR\tilde{M}_mH \right\|_2^2\\
%\end{align*}
Secondly, for $J_2$, noting equations (\ref{eq6}), (\ref{eq6a}) and (\ref{eq2aa}), the following equation holds
\begin{align*}
{J_2} =& \left\| {N_m(\tilde{Y}-R\tilde{M}){\lambda ^{ - 1}}{\sigma_n} } \right\|_2^2\\
&+ \left\| {N_m(\tilde{Y}-R\tilde{M}){\lambda ^{ - 1}}f^{(m)}{\sigma_q} } \right\|_2^2\\
=& \left\| {N_m(\tilde{Y}\tilde{B}^{-1}-R\tilde{M}_m)H} \right\|_2^2.
\end{align*}
where $H=\lambda^{-1}({\sigma_n}^2+|f^{(m)}|^2{\sigma_q}^2)^{1/2}$.\par
\par
Using Lemma 1, we can obtain
\begin{align*}
{J_2}=&\Bigg\|\sum\limits_{i = 1}^{n_p} {N_{m}}({p_i})\tilde Y({p_i}){H}({p_i})\tilde{B}_{net, Ii}^{-1}(\tilde{B}^{-1}_i-\tilde{B}^{-1}_i(\infty))\\
&\times\tilde{B}_{net, Oi}^{-1}+R_1-N_mR\tilde{M}_m H\Bigg\|_2^2
\end{align*}
where
\begin{align*}
&R_1=S+\sum\limits_{i = 1}^{n_p} {N_{m}}({p_i})\tilde Y({p_i}){H}({p_i})\tilde{B}_{net, Ii}^{-1}\tilde{B}^{-1}_i(\infty)\tilde{B}_{net, Oi}^{-1},\\
&\tilde B_{{net},Ii}=\tilde B_{{net},i - 1}({p_i}), \tilde B_{{net}, i-2}({p_i}),  \cdots \tilde B_{{net}, 1}({p_i}),\\
&\tilde B_{{net},Oi}=\tilde B_{net, {n_p}}({p_i}), \tilde B_{net, {n_p-1}}({p_i}), \cdots, \tilde B_{net, i + 1}({p_i}).
\end{align*}
Since
\begin{align*}
&S\in{{\rm \bf R}\mathcal{H}_\infty},\\
&\sum\limits_{i = 1}^{n_p} {N_{m}}({p_i})\tilde Y({p_i}){H}({p_i})\tilde{B}_{net, Ii}^{-1}\tilde{B}^{-1}_i(\infty)\tilde{B}_{net, Oi}^{-1}\in{{\rm \bf R}\mathcal{H}_\infty},
\end{align*}
then $R_1\in{{\rm \bf R}\mathcal{H}_\infty}$. Therefore, we can design
$
R=N_{m}^{\dag}R_1{H^{-1}\tilde M}_m^{-1},
$
then, we can obtain
\begin{align*}
{J_2^*}=&\Bigg\| \sum\limits_{i = 1}^{n_p} {N_{m}}({p_i})\tilde Y({p_i}){H}({p_i})\tilde{B}_{net, Ii}^{-1}(\tilde{B}^{-1}_i-\tilde{B}^{-1}_i(\infty)) \\
&\times\tilde{B}_{net, Oi}^{-1}\Bigg\|_2^2\\
=&\Bigg\| \sum\limits_{i = 1}^{n_p} {N_{m}}({p_i})\tilde Y({p_i}){H}({p_i})\tilde{B}_{net, Ii}^{-1}
\frac{2 {\rm {Re}}(p_i)}{s-p_i}\tilde{B}_{net, Oi}^{-1}\Bigg\|_2^2.
\end{align*}

\par
By Bezout identity (\ref{eq1}) and equations (\ref{eq7aa})(\ref{eq4}), we can get
\begin{align*}
{N_{m}}({p_i})Y({p_i})=&{N_{m}}({p_i})N_F^{-1}({p_i})\\
=&L^{-1}({p_i})L_F^{-1}({p_i})(f^{(m)}(p_i))^{-1}\\
=&\prod_{k=1}^{n_z}\frac{p_i+\bar z_k}{p_i-z_k}
\prod_{k=1}^{n_f}\frac{p_i+\bar s_k}{p_i-s_k}(f^{(m)}(p_i))^{-1}.
\end{align*}
\par
Then, we can obtain
\begin{align*}
J_2^*=&\Bigg\| \sum\limits_{i = 1}^{n_p}
\Bigg(\prod_{k=1}^{n_z}\frac{p_i+\bar z_k}{p_i-z_k}
\prod_{k=1,k\neq{}i}^{n_p}\frac{p_i+\bar p_k}{p_i-p_k}\Bigg)
\frac{{2{\mathop{\rm Re}} ({p_i})}}{{s - {p_i}}}\\
&\times\Bigg((f^{(m)}(p_i))^{-1}{H}({p_i})\prod_{k=1}^{n_f}\frac{p_i+\bar s_k}{p_i-s_k}\Bigg)\Bigg\|_2^2
\\
=&\sum\limits_{i,j = 1}^{n_p}{W}_{sys}^{i,j}{W}_{net}^{i,j},
\end{align*}
where
\begin{align*}
W_{sys}^{i,j}=&\frac{{4{\mathop{\rm Re}} ({p_i}){\mathop{\rm Re}} ({p_j})}}{{{{\bar p}_i} + {p_j}}}
\prod_{k=1}^{n_z}\frac{\bar p_i+ z_k}{\bar p_i-\bar z_k}
\prod_{k=1,k\neq{}i}^{n_p}\frac{\bar p_i+p_k}{\bar p_i-\bar p_k}\\
&\times\prod_{k=1}^{n_z}\frac{p_j+\bar z_k}{p_j-z_k}
\prod_{k=1,k\neq{}j}^{n_p}\frac{p_j+\bar p_k}{p_j-p_k},\\
W_{net}^{i,j}=&\lambda^{-2}\Big((\bar{f}^{m}(p_i))^{-1}\sigma_n+f^{m}(p_i)\sigma_q\Big)\\
&\times\Big((f^{m}(p_j))^{-1}\sigma_n+\bar{f}^{m}(p_j)\sigma_q\Big)\\
&\times\prod_{k=1}^{n_f}\frac{\bar p_i+ s_k}{\bar p_i-\bar s_k}
\prod_{k=1}^{n_f}\frac{p_j+\bar s_k}{p_j-s_k},
\end{align*}
and $W_{sys}^{i,j}$ is called the performance plant part, $W_{net}^{i,j}$ is called the performance network part.\par
In order to analyze the system performance more conveniently, we denote
\begin{align*}
%\bm g_j&=\prod_{k=1}^{n_z}\frac{p_j+\bar z_k}{p_j-z_k}\prod_{k=1,k\neq{}j}^{n_p}\frac{p_j+\bar p_k}{p_j-p_k},\\
\bm f_{net}(p_j)&=\lambda^{-1}\Big((f^{m}(p_j))^{-1}\sigma_n+\bar{f}^{m}(p_j)\sigma_q\Big)\prod_{k=1}^{n_f}\frac{p_j+\bar s_k}{p_j-s_k},
\end{align*}
%Then
%\begin{align*}
%J_2^*=&\frac{{4{\mathop{\rm Re}} ({p_i}){\mathop{\rm Re}} ({p_j})}}{{{{\bar p}_i} + {p_j}}}\bar{\bm g}_i\bm g_j\bar{\bm f}_i\bm f_j
%\end{align*}
Obviously, $W_{sys}^{i,j}=\bar{\bm f}_{net}(p_j)\bm f_{net}(p_j)$, we call $\bm f_{net}(\cdot)$ the network factor.
This completes the proof.
\end{IEEEproof}
\begin{remark}
Theorem \ref{thm1} clearly demonstrates that the quantitative relationship between tracking performance limitations and system characteristics (NMP zeros and unstable poles), network constraint characteristics (the power of communication noises and quantization noises, the NMP zeros of bandwidth and the MP part of bandwidth), the power of reference noises.
%And we give a separate form between the the network performance part and the performance plant part. This is more conducive to us to observe the impact of network constraints on tracking performance limitations.
\end{remark}
\begin{remark}
Theorem \ref{thm1} also demonstrates that we can completely divide the tracking performance limitation $J_2^*$ part of SISO NCSs into two parts: the plant part $W_{sys}^{i,j}$ and the network part $W_{sys}$. Moreover, $J_2^*$ part is independent of network constraints. If we want to detect the influences of various network constraints on the tracking performance limitation of the SISO NCSs, we just need to explore the relationship between the $W_{net}^{i,j}$ (or $\bm f_{net}(\cdot)$) and every network constraints. Moreover, there are some interesting results. Firstly, if any of the NMP zeros of bandwidth model is very close to one of the system's NMP zeros, the tracking performance limitation will be greatly deteriorated. Secondly, encoder-decoder and the NMP zeros of bandwidth will affect $W_{net}^{i,j}$ globally, while the MP part of bandwidth has the opposite effect between communication noise and quantization noise. This means that we can adjust the impact proportion of communication noise and quantization noise in performance limitations by the MP part of bandwidth, and this method is feasible in practice because bandwidth and encoder-decoder can be manually designed or intervened. In addition, the MP part of bandwidth and the NMP zeros of plant is coupling, this will make this method has more extensive features. If the bandwidth and quantization noise are not considered, this results can degraded to the situation in \cite{Chen16}.
\end{remark}
\begin{remark}
If there aren't the communication noise and the quantization noise of NCSs, the tracking performance limitation degenerate into $2{\sigma_r}^2\sum_{i = 1}^{{n_z}} {\mathop{\rm Re}} ({z_i}).$
This means that the effects of encoder-decoder and quantization noise can be offset by using two degrees of freedom control in the mean square sense. And, we will insight into the fact that there is the same result in MIMO case from the next section.
\end{remark}

\subsection{Tracking Performance Limitations of MIMO NCSs with Multiple Communication Constraints}

Similar to the proof in Subsection \ref{subse1}, it is easy to obtain
\begin{align}
J=&\mathcal{E}\left\{ {\left\|\bm y(s) - \bm r(s) \right\|_2^2 } \right\}\nonumber\\
=&\left\| {(I - NQ)\mathcal{U}} \right\|_2^2
+ \left\| {N(\tilde Y - R\tilde M_F){A^{ - 1}}\mathcal{V}} \right\|_2^2\nonumber\\
&+ \left\| {N(\tilde Y - R\tilde M_F){A^{ - 1}}F{\mathcal{Q}} } \right\|_2^2.\label{eq66}
\end{align}
Let
\begin{align}
J_1\triangleq&\left\| {(I - NQ)\mathcal{U}} \right\|_2^2,\label{eq8a}\\
J_2\triangleq&\left\| {N(\tilde Y - R\tilde M_F){A^{ - 1}}\mathcal{V}} \right\|_2^2\nonumber\\
&+ \left\| {N(\tilde Y - R\tilde M_F){A^{ - 1}}F{\mathcal{Q}} } \right\|_2^2.\label{eq9a}
\end{align}
\par
This following theorem gives us tracking performance limitations for the feedback configuration which is shown in Fig. 1. The MIMO plants is considered with distinct unstable poles and NMP zeros.
\begin{theorem}\label{thm2}
Considering the structure model as Fig. 1, and suppose that the reference noises $\bm r=[r_1, r_2,\ldots, r_m]^T$, the quantization noises $\bm q=[q_1, q_2,\ldots, q_m]^T$ as well as the communication noises $\bm n=[n_1, n_2,\ldots, n_m]^T$ should be zero mean i.i.d. GWN with PSD ${\sigma_r}_i^2, {\sigma_q}_i^2$ and ${\sigma_n}_i^2$, respectively. The model $F$ of the bandwidth restriction with $n_f$ distinct NMP zeros $s_k~(k = 1, \ldots ,n_f)$. It is assumed that $p_k~(k = 1, \ldots ,n_p )$ and $z_k~(k = 1, \ldots ,n_z )$ are unstable poles and NMP zeros of $G(s)$, respectively. Then,
\begin{align*}
 J^ * & = 2\sum\limits_{i = 1}^{{n_z}} {\mathop{\rm Re}} ({z_i})
\Psi_{r}\Psi_{\eta}\cos\angle ({\bm \upsilon_{r}, \check{\bm\eta}_i})\\
&+\sum\limits_{i,j = 1}^{n_p} \frac{{4{\mathop{\rm Re}\nolimits} ({p_i}){\mathop{\rm Re}\nolimits} ({p_j})}}{{{\bar p}_i}+ {p_j}}
{\bm h}_i^H{\bm f}_i^H{\bm f}_{j}{\bm h}_j
{\bm g}_j^H{\bm g}_i,
\end{align*}
where
\begin{align*}
&{\bm g}_i={\tilde B}_{net,Oi}^{ - H}{\bm \gamma_i},~~{\bm h}_j={\tilde B}_{net,Ij}^{ - 1}{\bm \gamma_j},\\
&{\bm f}_j=\prod_{k=1}^{n_z}\Big(I+\frac{2{\rm Re}z_k}{p_j+z_k}\eta_j\eta_j^H\Big){\bm f}_{net,j},\\
&{\bm f}_{net}(p_j)=F^{-1}(p_j)H(p_j).
\end{align*}
\end{theorem}
\begin{IEEEproof}
With the use of the equations (\ref{eq66}) as well as (\ref{eq8a})-(\ref{eq9a}) and referring that the parameter matrices $Q$ and $R$ are mutually independent, we have
\begin{align*}
J^*&=\inf_{K\in \mathcal{K}}J\\
&=\inf_{Q\in {\rm \bf R\mathcal{H}}_\infty}J_1+\inf_{R\in {\rm \bf R\mathcal{H}}_\infty}J_2\\
&=J_1^*+J_2^*.
\end{align*}
\par
Firstly, for $J_1$, we have
\begin{align*}
J_1^*=&\inf_{Q\in {\rm \bf R\mathcal{H}}_\infty}\left\| {(I - NQ)\mathcal{U}} \right\|_2^2\nonumber\\
=&\inf_{Q\in {\rm \bf R\mathcal{H}}_\infty}\left\| \Big(({L^{-1}-I)+(I - N_{m}Q)\Big)\mathcal{U}} \right\|_2^2\nonumber
\end{align*}
Owning to $L^{-1}(0)=I$ and $(L^{-1}-I)\mathcal{U}\in{\mathcal{H}_2^\perp}$, and $Q$ can be select so that
$(I - N(0)Q(0))\mathcal{U}=0$, we have $(I - N_{m}Q)\mathcal{U}\in{\mathcal{H}_2}$, then
\begin{align*}
J_1^*=&\left\| (L^{-1}-I)\mathcal{U}\right\|_2^2+\inf_{Q\in {\rm \bf R\mathcal{H}}_\infty}\left\|(I - N_{m}Q)\mathcal{U} \right\|_2^2\nonumber
\end{align*}
Because $N_m$ is an outer matrix function, we have
$$\inf_{Q\in {\rm \bf R\mathcal{H}}_\infty}\left\|(I - N_{m}Q)\mathcal{U} \right\|_2^2=0.$$
Thus,
\begin{align}
J_1^*=&\left\| (L^{-1}-I)\mathcal{U}\right\|_2^2\nonumber\\
=&\sum\limits_{i = 1}^{{n_z}} \left\| (I-L_i)\mathcal{U}\right\|_2^2\nonumber\\
=&\sum\limits_{i = 1}^{{n_z}}\left\| \frac{2{\rm Re}(z_i)}{s+\bar{z_i}}\bm \eta_i\bm \eta_i^H\mathcal{U}\right\|_2^2\nonumber\\
=&2\sum\limits_{i = 1}^{{n_z}} {\mathop{\rm Re}} ({z_i})\sum\limits_{j = 1}^{m}\sigma_{rj}^2{\bm \eta}_{ij}^2\nonumber\\
=&2\sum\limits_{i = 1}^{{n_z}} {\mathop{\rm Re}} ({z_i})
\Psi_{r}\Psi_{\eta_i}\cos\angle ({\bm \upsilon_{r}, \check{\bm\eta}_i}),\label{eq8b}
\end{align}
where $\Psi_{r}$ and $\Psi_{\eta_i}$ are reference noise PD-SSB and zero DSSB, ${\bm \upsilon_{r}}$ and ${\check{\bm\eta}_i}$ are reference noise PD-direction and zero transform direction.
\par
Secondly, for $J_2$, the following equation holds
\begin{align}
{J_2} =& \left\| N_{m}(\tilde Y - R\tilde M_F)H \right\|_2^2,\label{eq9b}
\end{align}
where
\begin{align}\label{eq18a}
H(s)&=\emph{diag}\{h_i(s)\},(i=1,2,\cdots,m),\\
h_i(s)&=\lambda_i^{-1}(\sigma_{ni}^2+|f_i(s)|^2\sigma_{qi}^2)^{1/2},\nonumber
\end{align}
then, we have
\begin{eqnarray}\label{eq10a}
{J_2} = \left\|N_{m}{\tilde Y}H - N_{m}R{\tilde M_F}H \right\|_2^2.
\end{eqnarray}
Furthermore, noting $H(s)$ is a diagonal, MP and stable transfer matrix, then ${\tilde M_F}(s)H(s)$ and ${\tilde M_F}(s)$ will have the same NMP zeros, but the corresponding NMP zero direction will be changed by the networked constraints. Thus, we can perform an all-pass factorization
\begin{align}\label{eq11}
{\tilde M}_F(s)H(s)={\tilde M}_{net}(s)\tilde{B}_{net}(s),
\end{align}
where $\tilde{B}_{net}(s)$ is the allpass factor and ${\tilde M}_{net}(s)$ is the MP part of ${\tilde M}_F(s)H(s)$.
\par
We may construct $B_{net}(s)$ as
\begin{align}\label{eq12}
 \tilde{B}_{net}(s)=&\prod_{i=1}^{n_p}\tilde{B}_{{\rm{net}},i}(s),
 {\tilde{B}_{net,i}}(s):=I-\frac{2{\rm Re}(p_i)}{s+\bar{p}_i}\bm \gamma_i\bm \gamma_i^H,
\end{align}
The unitary vector $\bm \gamma_i=[\gamma_{i1},\cdots,\gamma_{im}]^T$ depends on the network factor $H(s)$.
\par
Then
\begin{eqnarray}
{J_2} = & \Big\| {{N_{\rm{m}}}\tilde YH\tilde B_{net}^{ - 1} - {N_{\rm{m}}}R{{\tilde M}_{net}}} \Big\|_2^2.\label{eq13}
\end{eqnarray}
\par
According to Lemma \ref{le1}, it is easily obtained that
\begin{align*}
{N_{\rm{m}}}\tilde YH\tilde B_{net} ^{ - 1}
=&S + \sum\limits_{i = 1}^{n_p} {N_{\rm{m}}}({p_i})\tilde Y({p_i}){H}({p_i})\\
&\times\tilde B_{{net},I}^{ - 1}({p_i})\tilde B_{net,i}^{ - 1}(s)\tilde B_{{net},O}^{-1}(p_i),
\end{align*}
where $S\in{}{\rm \bf R\mathcal{H}}_\infty$ and
\begin{align*}
\tilde B_{{net},Ii}&=\tilde B_{{net},i - 1}({p_i}), \tilde B_{{net}, i-2}({p_i}),  \cdots \tilde B_{{net}, 1}({p_i}),\\
\tilde B_{{net},Oi}&=\tilde B_{net, {n_p}}({p_i}), \tilde B_{net, {n_p-1}}({p_i}), \cdots, \tilde B_{net, i + 1}({p_i}).
\end{align*}
Therefore, we have
\begin{align}\label{eq6aa}
{J_2} =& \Big\|S + \sum\limits_{i = 1}^{n_p} {N_{\rm{m}}}({p_i})\tilde Y({p_i}){H}({p_i})
\tilde B_{net, Ii}^{ - 1}\Big(\tilde B_{net,i}^{ - 1}(s)\nonumber\\
& - \tilde B_{net, i}^{ - 1}(\infty )\Big)\tilde B_{net, Oi}^{ - 1}+ \sum\limits_{i = 1}^{Ns} N_{\rm{m}}(p_i)\tilde Y(p_i){H}({p_i})\nonumber\\
&\times\tilde B_{net, Ii}^{ - 1}\tilde B_{net, i}^{ - 1}(\infty )\tilde B_{net, Oi}^{ - 1} - {N_{\rm{m}}}R{{\tilde M}_{net}}\Big\|_2^2\nonumber\\
=& \Big\|R_1 + \sum\limits_{i = 1}^{n_p} {N_{\rm{m}}}({p_i})\tilde Y({p_i})H({p_i})
\tilde B_{net, Ii}^{ - 1}\Big(\tilde B_{net,i}^{ - 1}(s)\nonumber\\
& - \tilde B_{ net, i}^{ - 1}(\infty )\Big)\tilde B_{net, Oi}^{ - 1}- {N_{\rm{m}}}R{{\tilde M}_{net}}\Big\|_2^2,
\end{align}
where
\begin{align}\label{eq7a}
R_1=& S+\sum\limits_{i = 1}^{n_p} {N_{\rm{m}}}({p_i})\tilde Y({p_i})
H({p_i})\tilde B_{net, I}^{ - 1}({p_i})\nonumber\\
&\times\tilde B_{net, i}^{ - 1}(\infty )\tilde B_{net, O}^{ - 1}({p_i}).
\end{align}
\par
Noting
\begin{align*}
\sum\limits_{i = 1}^{n_p} {N_{\rm{m}}}({p_i})\tilde Y({p_i})&H({p_i})
\tilde B_{net, Ii}^{ - 1}\Big(\tilde B_{net,i}^{ - 1}(s)\nonumber\\
& - \tilde B_{ net, i}^{ - 1}(\infty )\Big)\tilde B_{net, Oi}^{ - 1}\in{\mathcal{H}_2^\bot}
\end{align*}
Then, from equation (\ref{eq6aa}), we can get
\begin{align*}
J_2^* =& \mathop {\inf }\limits_{R \in {\rm \bf R\mathcal{H}}_\infty}\Bigg\|\sum\limits_{i = 1}^{n_p} {N_{\rm{m}}}({p_i})\tilde Y({p_i})H(p_i)\tilde{B}_{net, Ii}^{-1}(\tilde B_{net, i}^{ - 1}(s)\\
&- \tilde B_{net, i}^{ - 1}(\infty ))\tilde{B}_{net, Oi}^{-1} + R_1 - {N_{\rm{m}}}R{{\tilde M}_{net}}\Big\|_2^2\\
= &\Bigg\| \sum\limits_{i = 1}^{n_p} {N_{\rm{m}}}({p_i})\tilde Y({p_i})H(p_i)\tilde{B}_{net, Ii}^{-1}
(\tilde B_{net, i}^{ - 1}(s)\\
& - \tilde B_{net, i}^{ - 1}(\infty ))\tilde{B}_{net, Oi}^{-1}  \Bigg\|_2^2\\
&+ \mathop {\inf }\limits_{R \in {\rm \bf R\mathcal{H}}_\infty} \Bigg\| R_1 - {N_{\rm{m}}}R{{\tilde M}_{ net}} \Bigg\|_2^2.
\end{align*}
\par
Noting equation (\ref{eq7a}), we have
\begin{align*}
\sum\limits_{i = 1}^{n_p} {N_{\rm{m}}}({p_i})&\tilde Y({p_i})H(p_i)\tilde{B}_{net, Ii}^{-1}\\
&\times\tilde B_{net, i}^{ - 1}(\infty )\tilde{B}_{net, Oi}^{-1}\in{\rm \bf R\mathcal{H}}_\infty,
\end{align*}
and $ S\in{}{\rm \bf R\mathcal{H}}_\infty.$
This means that $R_1\in {{\rm \bf R\mathcal{H}}_\infty}$.
And $N_{m}$ is right reversible, $M_{net}$ is reversible. Therefore, we can design
\begin{align*}
R=N_{m}^{^\dag}R_1{\tilde M}_{net}^{-1},
\end{align*}
where $N_{m}^\dag$ is right inverse of $N_{m}$.
Accordingly, the optimal performance
\begin{align}
J_2^*=& \Bigg\| \sum\limits_{i = 1}^{n_p} {N_{\rm{m}}}({p_i})\tilde Y({p_i}){H}({p_i})\tilde{B}_{net, Ii}^{-1}
(\tilde B_{net, i}^{ - 1}(s)\nonumber\\
& - \tilde B_{net, i}^{ - 1}(\infty ))\tilde{B}_{net, Oi}^{-1}\Bigg\|_2^2\nonumber\\
=& \Bigg\| \sum\limits_{i = 1}^{n_p} {N_{\rm{m}}}({p_i})\tilde Y({p_i}){H}({p_i})\tilde{B}_{net, Ii}^{-1}
\frac{{2{\mathop{\rm Re}\nolimits} ({p_i})}}{{s - {p_i}}}\nonumber\\
&\times{\gamma _i}\gamma _i^H\tilde{B}_{net, Oi}^{-1}  \Bigg\|_2^2.\label{eq15b}
\end{align}
\par
Noting the double Bezout identity (\ref{eq1}), it yields that
\begin{align*}
I&={\tilde X}(p_j)M_F(p_j)-{\tilde Y}(p_j)N_F(p_j)\\
&=-{\tilde Y}(p_j)N_F(p_j).
\end{align*}
\par
Then, we can obtain
$${\tilde Y}(p_j)=-N_{m}^\dag(p_j)L^{-1}(p_j)F^{-1}(p_j)$$
Moreover, noting equation (\ref{eq4}), and $F$ is diagonal matrix.
Therefore
\begin{align}
J_2^*=& \sum\limits_{i,j = 1}^{n_p} \frac{{4{\mathop{\rm Re}\nolimits} ({p_i}){\mathop{\rm Re}\nolimits} ({p_j})}}{{{\bar p}_i}+{p_j}}
\bm \gamma _i^H\tilde{B}_{net, Ii}^{-H}H^H({p_i}){{\tilde Y}^H}({p_i})N_{m}^H({p_i})\nonumber\\
&\times {N_{\rm{m}}}({p_j})\tilde Y({p_j}){H}({p_j})\tilde{B}_{net, Ij}^{-1}{\bm \gamma _j}
\bm \gamma _j^H\tilde{B}_{net, Oj}^{-1}\tilde{B}_{net, Oi}^{-H}{\bm \gamma _i}\nonumber\\
=& \sum\limits_{i,j = 1}^{n_p} \frac{{4{\mathop{\rm Re}\nolimits} ({p_i}){\mathop{\rm Re}\nolimits} ({p_j})}}{{{\bar p}_i}+{p_j}}
\bm \gamma _i^H\tilde{B}_{net, Ii}^{-H}{\bm f}^H_{net}(p_i)\nonumber\\
&\times {L^{-H}(p_i)}{L^{-1}(p_j)}{\bm f}_{net}(p_j)\tilde{B}_{net, Ij}^{-1}{\bm \gamma _j}\nonumber\\
&\times \bm \gamma _j^H\tilde{B}_{net, Oj}^{-1}\tilde{B}_{net, Oi}^{-H}{\bm \gamma _i}\label{eq24}
\end{align}
where
\begin{align}\label{eq29}
&{\bm f}_{net}(p_j)=F^{-1}(p_j)H(p_j)
%=&L_f^{-1}(p_j)\emph{diag}\{\lambda_k^{-1}(\bar{f}_k^{(m)}(p_j)\sigma_{nk}^2+f_k^{(m)}(p_j)\sigma_{qk}^2)^{1/2}\}
\end{align}
and we define ${\bm f}_{net}(\cdot)$ as the network factor.
This completes the proof.
\end{IEEEproof}
\begin{remark}
Different from the SISO systems, we can find out the tracking performance limitations of MIMO systems have some new features from theorem \ref{thm2}. Firstly, the network constraints and the system characteristics are more closely coupled. However, the network factor ${\bm f}_{net}$ can completely contain the impact of all network constraints ${\bm f}_{net}$ apart from the effects of bandwidth on the zero directions of plant. Secondly, the tracking performance limitations are uniquely determined by the NMP zeros and unstable poles of the plant, direction of NMP zeros and unstable poles, and network constraints. In addition, ${\bm f}_{net}$ demonstrates that this performance limitations depends not only on the power distribution of the reference noises, communication noise and quantization noise, and the allocation of bandwidth and encoder-decoder, but also on NMP zeros of bandwidth and MP part of bandwidth.
\end{remark}
\medskip
\begin{remark}
The performance $J_1^*$ part in theorem \ref{thm2} also demonstrates that these angles between the reference noise PD-Directions and each NMP zero direction will affect the tracking performance limitations. And, these varies from $0^\circ$ to $90^\circ$. If the transform NMP zeros directions and the reference noises power distribution are all orthogonal (i.e., their angles $\angle(\upsilon_r,\bm \check{\eta}_i)=90^\circ$), the NMP zero and reference signal will not have effects on the tracking performance limitations of NCSs. Additionally, if the plant is MP, the reference signal tracking won't be interfering with the performance limitations. If our communication constraints only consider communication noises, our results can degenerate to similar results in \cite{10Ding}.
\end{remark}
\begin{remark}
As can be seen from $\gamma_i$ in theorem \ref{thm2}, bandwidth allocation will affect each directions of unstable poles.  In addition, from (\ref{eq18a})(\ref{eq29}), we have
${\bm{f}}_{net}(p_j)
=L_f^{-1}(p_j)\emph{diag}\{\lambda_k^{-1}(\bar{f}_k^{(m)}(p_j)\sigma_{nk}^2+f_k^{(m)}(p_j)\sigma_{qk}^2)^{\frac{1}{2}}\}.
$
This equation demonstrates that the allocations of MP part of bandwidth and encoder-decoder in each channel will affect corresponding channel of network factor, but UMP zero part of bandwidth may be affect all of channel. In addition, bandwidth also affect each directions of unstable poles. Moreover, since bandwidth and encode-decode can be adjusted by manual intervention, it is a feasible method allocate different channel performance of the system by adjust the allocation of bandwidth and encode decode. This may be a viable direction for performance allocation research.
%\begin{align*}
%{\bm f}_{net}(p_j)= &L_f^{-1}(p_j)\emph{diag}\{\lambda_k^{-1}(\bar{f}_k^{(m)}(p_j)\sigma_{nk}^2\\
%+f_k^{(m)}(p_j)\sigma_{qk}^2)^{1/2}\},\\
%= &\Big(\frac{p_j+\bar{s}_j}{p_j+\bar{s}_j}\Big)\eta_{fi}\eta_{fi}^H+I\\
%&\times\emph{diag}\Big[\lambda_1^{-1}(\sigma_{n1}^2\bar{f}_1^{(m)}(p_j)+f_1^{(m)}(p_j)\sigma_{q1}^2)^{1/2},\\
%&~~~~~~~~~~\lambda_2^{-1}(\sigma_{n2}^2+f_2^{(m)}(p_j)\sigma_{q2}^2)^{1/2}, \cdots, \\ %&~~~~~~~~~~~~~~~~\lambda_m^{-1}(\sigma_{nm}^2+f_m^{(m)}(p_j)\sigma_{qm}^2)^{1/2}\Big]
%\end{align*}
\end{remark}
%\begin{center}
%\includegraphics[width=8.5cm]{SISO.eps}
%\includegraphics[height=3.1cm,width=7.6cm]{Fig.1.pdf}
%{\footnotesize \centerline{{\bf Fig. 2.} The SISO Plant Control Scheme}}
%\end{center}
\vspace{5pt}
\par
Furthermore, we analyze the circumstances where multiple unstable poles may exist with parallel or orthogonal directions.
\begin{corollary}\label{co1}
As Fig. 1, the NCS considered here which is based on the structural model is demonstrated. Suppose that $G(s)$ is minimum, and $z$ is the NMP zero. The results are as follows.
\begin{description}
  \item[(a)] Each directions of unstable pole of $G(s)$ are parallel with unstable pole direction $\omega$, the tracking performance limitation is
%\begin{align*}
%J^*=&\sum\limits_{i,j = 1}^{n_p} \frac{{4{\mathop{\rm Re}\nolimits} ({p_i}){\mathop{\rm Re}\nolimits} ({p_j})}}{{{\bar p}_i}+ {p_j}}\frac{1}{(\|H_i^{-1}\hat{{\bm \omega}}\|^2\|H_j^{-1}\hat{{\bm \omega}}\|^2)}\\
%&\Big(\hat{{\bm \omega}}^HF_i^{-H}F_j^{-1}\hat{{\bm \omega}}+\bar{b}_{Ii}\hat{{\bm \omega}}^HH_i^{-H}\hat{{\bm \omega}}\hat{{\bm \omega}}^H{\bm f}_{net,i}^{H}F_j^{-1}\hat{{\bm \omega}}\\
%&+b_{Ij}\hat{{\bm \omega}}^HF_i^{-H}{\bm f}_{net,j}\hat{{\bm \omega}}\hat{{\bm \omega}}^HH_j^{-H}\hat{{\bm \omega}}\\
%&+\bar{b}_{Ii}b_{Ij}\hat{{\bm \omega}}^HH_i^{-H}\hat{{\bm \omega}}\hat{{\bm \omega}}^H{\bm f}_{net,i}^{H}{\bm f}_{net,j}\hat{{\bm \omega}}\hat{{\bm \omega}}^HH_j^{-1}\hat{{\bm \omega}}\Big)\\
%&\times\Big((b_{Oi}+b_{Oj}+b_{Oi}b_{Oj})\hat{{\bm \omega}}^HH_i^{-H}\hat{{\bm \omega}}
%\hat{{\bm \omega}}^HH_j^{-1}\hat{{\bm \omega}}\\
%&+\hat{{\bm \omega}}^HH_i^{-H}H_j^{-1}\hat{{\bm \omega}}\Big)
%\end{align*}
\begin{align*}
J^*=&\sum\limits_{i,j = 1}^{n_p} \frac{{4{\mathop{\rm Re}\nolimits} ({p_i}){\mathop{\rm Re}\nolimits} (p_j)}}{({\bar p}_i+ {p_j})\|H^{-1}(p_i)\hat{\bm \omega}\|^2\|H^{-1}(p_j)\hat{\bm \omega}\|^2}\\
&\times\hat{\bm\omega} _i^HH_i^{-H}\tilde{B}_{net, Ii}^{-H}{\bm f}^H_{net,i}{\bm f}_{net,j}\tilde{B}_{net, Ij}^{-1}H_j^{-1}{ \hat{\bm\omega} _j}\nonumber\\
&\times\hat{\bm \omega} _j^HH_j^{-1}\tilde{B}_{net, Oj}^{-1}\tilde{B}_{net, Oi}^{-H}H_i^{-1}{ \hat{\bm\omega} _i}
\end{align*}
where
\begin{align*}
\tilde{B}_{net, Ii}^{-1}=& \left[\prod_{k=1}^{i-1}\frac{p_i+\bar{p}_k}{p_i-{p}_k}-1\right]\hat{{\bm \omega}}\hat{{\bm \omega}}^H+I,\\
\tilde{B}_{net, Oi}^{-1}=& \left[\prod_{k=i+1}^{n_p}\frac{p_i+\bar{p}_k}{p_i-{p}_k}-1\right]\hat{{\bm \omega}}\hat{{\bm \omega}}^H+I.
\end{align*}
  \item[(b)] Each directions of unstable pole of $G(s)$ are orthogonal, Suppose that the the directions are $\omega_k~(k = 1, \ldots ,n_p)$, then
\begin{align*}
J^*=&
\sum\limits_{i = 1}^{n_p}\frac{ 2{\mathop{\rm Re}\nolimits} ({p_i})}{\|H^{-1}(p_i)\hat{\bm \omega}\|^4}\\
&\times\hat{\bm\omega} _i^HH_i^{-H}\tilde{B}_{net, Ii}^{-H}{\bm f}^H_{net,i}{\bm f}_{net,i}\tilde{B}_{net, Ii}^{-1}H_i^{-1}{ \hat{\bm\omega} _i}\nonumber\\
&\times\hat{\bm \omega} _i^HH_i^{-1}\tilde{B}_{net, Oi}^{-1}\tilde{B}_{net, Oi}^{-H}H_i^{-1}{ \hat{\bm\omega} _i}
\end{align*}
where
\begin{align*}
\tilde{B}_{net, Ii}^{-1}=& \sum_{k=1}^{i-1}\frac{2{\rm Re}{p_i}}{p_i-{p}_k}\hat{{\bm \omega}}_k\hat{{\bm \omega}}_k^H+I,\\
\tilde{B}_{net, Oi}^{-1}=& \sum_{k=i+1}^{n_p}\frac{2{\rm Re}{p_i}}{p_i-{p}_k}\hat{{\bm \omega}}_k\hat{{\bm \omega}}_k^H+I.
\end{align*}
\end{description}

\end{corollary}

\begin{IEEEproof}
Let
\begin{align*}
\bm \gamma_i=&\frac{H^{-1}\hat{\bm \omega}}{\|H^{-1}\hat{\bm \omega}\|}.
\end{align*}
From the theorem $\ref{thm2}$ and noting equation (\ref{eq12}), it is easy to prove.
\end{IEEEproof}
\begin{remark}
Corollary \ref{co1} demonstrates that different zero directions will result in different system performance, this property is similar to the classic control systems \cite{chen2000limitations,freudenberg2003fundamental,middleton2004tracking}.
\end{remark}
\par
For the purpose of providing more conceptual guidance into this result, we take into consideration that there is one single unstable pole $p$ with steering vector $\bm \omega$ as well as a simple NMP zero $z$ with steering vector $\bm \eta$ in the MP plant. In Theorem 1, we can get the corollary as follows.
%\begin{corollary}\label{cor1a}
%Suppose that $p$ is the unstable pole of $G(s)$, and $z$ is the NMP zero. Then the system in Fig. $\ref{fig1a}$ have the tracking performance limitations as follows.
%\begin{align*}
%J^ * = & 2{\mathop{\rm Re}} ({z})\Phi_{r}\cos\angle ({\bm \eta },{\bm v}_{r})\\
% &+2{\rm Re}(p)\bm \gamma^H[\mathcal{V}+\mathcal{Q}F^H_m(p)]A^{-1}F^{-H}(p)L^H(p)\\
%&\times L(p)F^{-1}(p)A^{-1}(\mathcal{V}+\mathcal{Q}F^H_m(p))\bm \gamma
%\end{align*}
%\end{corollary}
%\begin{IEEEproof}
%From the equation (\ref{eq24}), we have
%\begin{align*}
%J_2^ * = & 2{\rm Re} (p)\bm \gamma^H\bm f_{net,1}^HL^{-H}(p)L^{-1}(p)\bm f_{net,1}^H\bm \gamma\\
%= &2{\rm Re} (p)\bm \gamma^H\bm f_{net,1}^H\Big(I+\frac{2{\rm Re}(z)}{\bar{p}-\bar{z}}\eta\eta^H\Big)\\
%&\times\Big(I+\frac{2{\rm Re}(z)}{p-z}\eta\eta^H\Big)\bm f_{net,1}^H\bm \gamma\\
%=&2{\rm Re} (p)\bm \gamma^H\bm f_{net,1}^H\bm f_{net,1}\bm \gamma\\
%&+\frac{8{\rm Re} (p){\rm Re}^2 (z)}{|p-z|^2}\bm \gamma^H\bm f_{net,1}^H\eta_i\eta_i^H
%\bm f_{net,1}\bm \gamma
%\end{align*}
%Let
%\begin{align*}
%\bm \gamma=\frac{\bm f_{net,1}^{-1}\hat{\bm \omega}}{\|\bm f_{net,1}\hat{\bm \omega}\|}
%\end{align*}
%\end{IEEEproof}
\begin{corollary}\label{cor1}
As Fig. 1, the NCS considered here which is based on the structural model is demonstrated. The reference signal $\bm r=[r_1, r_2,\ldots, r_m]^T$ with PSD ${\sigma_r}_i^2$. Suppose that $p$ is the unstable pole of MP plant $G(s)$, and $z$ is the NMP zero. The results are as follows.
\begin{description}
  \item[(a)] If the quantization noise is not considered, i.e., $\bm q=0$, the tracking performance limitations can be represented as follows
\begin{align*}
J^ * = &2{\rm Re}\{p\}\Psi_{F}(p)\Psi_{\check{\hat{\omega}}}\frac{\cos\angle({\bm \upsilon}_{F}(p),{\check{\hat{\bm \omega}}})}
{\|A\mathcal{V}^{-1}{ \hat{\bm\omega}}\|^2}.
\end{align*}
Specially, when every channel have the same bandwidth, then
\begin{align*}
J^*=&2{\rm Re}\{p\} |f^{(m)}(p)|^{-2}\\
&\times\prod_{k=1}^{n_f}\left|\frac{p+\bar{s}_k}{p-s_k}\right|^{2}\frac{\sum_{i=1}^m\cos\angle({\bm e}_i,{\bm \hat{\omega}})}{\|A\mathcal{V}^{-1}{ \hat{\bm\omega}}\|^2},
\end{align*}
  \item[(b)] If the communication noise is not considered, i.e., $\bm n=0$, the tracking performance limitations can be represented as follows
\begin{align*}
J^ * = &2{\rm Re}\{p\}\Psi_{F}(p)\Psi_{\check{\hat{\omega}}}\frac{\cos\angle(\bm\upsilon_{F}(p),
\check{\hat{\bm \omega}})}{\|A\mathcal{Q}^{-1}F_m^{-1}(p){\hat{\bm \omega}}\|^2}.
\end{align*}
  \item[(c)] If the bandwidth is unrestricted, i.e., $F=I$, the tracking performance limitations can be represented as follows
\begin{align*}
J^ * = &2{\rm Re}\{p\}\frac{\sum_{i=1}^m\cos\angle({\bm e}_i,{{\bm\omega}})}{\|H^{-1}{{\bm\omega}}\|^2}.
\end{align*}
  \item[(d)] If there are $\bm q=0, A=I, $,
  %i.e only the communication noises and bandwidth are considered for network's constraints, then the tracking performance limitations can be represented as follows
%\begin{align*}
%J^ * = & 2{\mathop{\rm Re}} ({z})
%\Psi_{r}\Psi_{\eta}\cos\angle ({\bm \upsilon_{r}, \check{\bm\eta}})\\
%&+2{\rm Re}\{p\}\Psi_{n_F}(p)\Psi_{\gamma}\cos\angle(\bm{v}_{n_F}(p),\bm\gamma).
%\end{align*}
and every channel have the same bandwidth, then
\begin{align*}
J^*=&2{\rm Re}\{p\}|f^{(m)}(p)|^{-2}\prod_{k=1}^{n_f}\left|\frac{p+\bar{s}_k}{p-s_k}\right|^{2}\\
&\times\Psi_n\Psi_{\check{\gamma}}\cos\angle(\bm{\upsilon}_n,{\check{\bm\gamma}}).
\end{align*}
  \item[(e)]
If there are $\bm n=0, A=I$, i.e only the quantization noise and bandwidth are considered for network's constraints, then
\begin{align*}
J^*= & 2{\rm Re}\{p\}\prod_{k=1}^{n_f}\left|\frac{p+\bar{s}_k}{p-s_k}\right|^{2}\Psi_q\Psi_{{\check{\gamma}}}\cos\angle(\bm{\upsilon}_q,\check{\bm\gamma}).
\end{align*}
\end{description}
\end{corollary}
\begin{IEEEproof}
From the theorem $\ref{thm2}$ and noting equation (\ref{eq12})(\ref{eq15b}),
\par
For (a), we can construct
\begin{align}\label{eq25a}
{\bm \gamma}=\frac{A\mathcal{V}^{-1}{\bm \hat{\omega}}}{\|A\mathcal{V}^{-1}{\bm \hat{\omega}}\|}.
\end{align}
\par
Then
\begin{align}
J_2^*=&2{\rm Re}\{p\}{\rm Tr}\Bigg\{\frac{{F^{-1}}(p){\bm \hat{\omega}}
{\bm \hat{\omega}}^H{{F}^{-H}}(p)}{\|A\mathcal{V}^{-1}
{\bm \hat{\omega}}\|^2}\Bigg\}\nonumber\\
=&2{\rm Re}\{p\}\Psi_{F}(p)\Psi_{\check{\hat{\omega}}}\frac{\cos\angle({\bm \upsilon}_{F}(p),{\check{\hat{\bm \omega}}})}
{\|A\mathcal{V}^{-1}{ \hat{\bm\omega}}\|^2},\label{eq16a}
\end{align}
where
${\bm \upsilon}_{F}(p)$ and ${\check{\hat{\bm \omega}}}$ are the bandwidth transform allocation direction and transform pole directions, respectively. $\Psi_{F}(p)$ and $\Psi_{\check{\hat{\omega}}}$ are TA-SSB and BI-DSSB, respectively.
\par
%If the bandwidth is the same on each channel, similar we can construct, then
%\begin{align}
%J_2^*=2{\rm Re}\{p\}|f(p)|^{-2}\Psi_q\Psi_{\gamma}\cos\angle(\bm{\upsilon}_q,\check{\bm\gamma})\end{align}
Specially, if the bandwidth is the same on each channel, then, we have
\begin{align}
J_2^*=&2f^{-2}(p){\rm Re}\{p\}{\rm Tr}\Bigg\{\frac{{\bm \hat{\omega}}{\bm \hat{\omega}}^H}{\|H^{-1}{ \hat{\bm\omega}}\|^2}\Bigg\}\nonumber\\
=&2{\rm Re}\{p\} |f^{(m)}(p)|^{-2}\prod_{k=1}^{n_f}\left|\frac{p+\bar{s}_k}{p-s_k}\right|^{2}\frac{\sum_{i=1}^m\cos\angle({\bm e}_i,{\bm \hat{\omega}})}{\|A\mathcal{V}^{-1}{ \hat{\bm\omega}}\|^2},
\end{align}
where ${\bm e}_i$ is Euclidean coordinates, where the $i$th element is equal to one and is
the only nonzero entry.
\par
For (b), similar we can construct
\begin{align*}
{\bm \gamma}=\frac{A\mathcal{Q}^{-1}F_m^{-1}{ \hat{\bm\omega}}}{\|A\mathcal{Q}^{-1}F_m^{-1}{ \hat{\bm\omega}}\|}.
\end{align*}
From the theorem $\ref{thm2}$ and noting equation (\ref{eq12}), we can obtain
\begin{align}
J_2^*=&2{\rm Re}\{p\}{\rm {Tr}}\Bigg\{\frac{{F^{-1}}(p){\hat{\bm\omega}}{ \hat{\bm\omega}}^H{{F}^{-H}}(p)}
{\|A\mathcal{Q}^{-1}F_m^{-1}(p){\hat{\bm\omega}}\|^2}\Bigg\}\nonumber\\
=&2{\rm Re}\{p\}\Psi_{F}(p)\Psi_{\check{\hat{\omega}}}\frac{\cos\angle(\bm\upsilon_{F}(p),
\check{\hat{\bm \omega}})}{\|A\mathcal{Q}^{-1}F_m^{-1}(p){\hat{\bm \omega}}\|^2}.\label{eq17}
\end{align}
\par
For (c), similar we can construct
\begin{align*}
{\bm \gamma}=\frac{H^{-1}{{\bm \omega}}}{\|H^{-1}{{\bm \omega}}\|}.
\end{align*}
From equations (\ref{eq12}), we can obtain
\begin{align}
J_2^*=&2{\rm Re}\{p\}{\rm Tr}\Bigg\{\frac{{{\bm\omega}}{{\bm\omega}}^H}{\|H^{-1}{{\bm\omega}}\|^2}\Bigg\}\nonumber\\
=&2{\rm Re}\{p\}\frac{\sum_{i=1}^m\cos\angle({\bm e}_i,{{\bm\omega}})}{\|H^{-1}{{\bm\omega}}\|^2},\label{eq18}
\end{align}
where
$$
H=\emph{diag}\{\lambda_i^{-1}(\sigma_{ni}^2+\sigma_{qi}^2)^{{1}/{2}}.
$$
%\begin{align*}
%J_2^*=& 2{\mathop{\rm Re}\nolimits} ({p})\bm \gamma^HH^H(p)H(p){\bm \gamma}\nonumber\\
%=&2 {\mathop{\rm Re}} ({z})\sum\limits_{j = 1}^{m}h_{j}^2{\bm \omega}_{j}^2\nonumber\\
%=&2 {\mathop{\rm Re}} ({z})\Psi_{h}\Psi_{\gamma}\cos\angle ({\bm \upsilon_{h}, \check{\bm\gamma}}),
%\end{align*}
%where
%\begin{align*}
%\Psi_{h}&=\|H(p)\|_F,\\
%{\bm \upsilon}_{h}&= [h_1(p),h_2(p),\cdots,h_m(p)]/\Phi_{h},
%\end{align*}
%$\Psi_{h}$ and $\Psi_{\omega_i}$ are called communication constraints coupling allocation SSB and zero DSSB, ${\bm \upsilon_{h}}$ and $\check{\bm\omega}$ are called communication constraints coupling %allocation direction and zero transform direction.
%where
%\begin{align*}
%H=&\emph{diag}\{\lambda_i^{-2}(\sigma_{ni}^2+\sigma_{qi}^2)^\frac{1}{2}\}.
%\end{align*}
\par
For (d), if there are $q=0, A=I$,
%\begin{align}
%J_2^*=&2\Big\|F^{-1}(p)V\frac{2{\rm Re}(p)}{s-p}{\bm\gamma \bm\gamma}^H\Big\|_2^2\nonumber\\
%=&2{\rm Re}\{p\}\Psi_{n_F(p)}\Psi_{\gamma}\cos\angle(\bm{v}_{n_F}(p),\bm\check{\gamma}),\label{eq19}
%\end{align}
%where
%\begin{align*}
%\Psi_{n_F}&=\|\upsilon_F\Psi_F+\upsilon_n\Psi_n\|_F,\\
%{\bm \upsilon}_{n_F}&= (\upsilon_F\Psi_F+\upsilon_n\Psi_n)/\Phi_{Fn},\\
%\Psi_{\gamma}&=\left\|[\gamma_{i1}^2,\gamma_{i2}^2,\dots,\gamma_{i1}^2]^T\right\|_F,\\
%{\bm \upsilon}_{\check{\gamma}}&=[\gamma_{i1}^2,\gamma_{i2}^2,\dots,\gamma_{i1}^2]^T/\Psi_{\gamma}.
%\end{align*}
%where ${\bm \upsilon}_{Fn}$ is called the bandwidth $F$ and communication noise $n$ coupling power distribution direction (BN-CPD-Direction), $\Psi_{n_F}$ is called $\Psi_{n_F}$ is the bandwidth and communication noise coupling square-sum block (BN-CDSSB).\par
and when every channel have the same bandwidth, then
\begin{align}
J_2^*=&2\Big\|F^{-1}(p)V\frac{2{\rm Re}(p)}{s-p}{\bm\gamma \bm\gamma}^H\Big\|_2^2\nonumber\\
=&2{\rm Re}\{p\}|f^{m}(p)|^{-2}\prod_{k=1}^{n_f}\left|\frac{p+\bar{s}_k}{p-s_k}\right|^{2}\Psi_n\Psi_{\check{\gamma}}\cos\angle(\bm{\upsilon}_n,\check{{\bm\gamma}}).\label{eq20}
\end{align}
where
\begin{align}
\Psi_{\check{\gamma}}&=\left\|[\gamma_{1}^2,\gamma_{2}^2,\dots,\gamma_{m}^2]^T\right\|_F,\\
\check{\bm\gamma}&=[\gamma_{1}^2,\gamma_{2}^2,\dots,\gamma_{m}^2]^T/\Psi_{\check{\gamma}}
\end{align}
 $\Psi_{\check{\gamma}}$ and $\check{\bm\gamma}$ are called zero transform bandwidth interference DSSB and zero transform bandwidth interference direction, respectively.
\par
For (e), if there are $\bm n=0, A=I$, and every channel have the same bandwidth, then
\begin{align}
J_2^*=&\Big\|F^{-1}(p)\mathcal{Q}F_m(p)\frac{2{\rm Re}(p)}{s-p}{\bm\gamma \bm\gamma}^H\Big\|_2^2\nonumber\\
=&\Big\|L^{-1}(p)\mathcal{Q}\frac{2{\rm Re}(p)}{s-p}{\bm\gamma \bm\gamma}^H\Big\|_2^2\nonumber\\
=&2{\rm Re}\{p\}\prod_{k=1}^{n_f}\left|\frac{p+\bar{s}_k}{p-s_k}\right|^{2}\Psi_q\Psi_{{\gamma}}\cos\angle(\bm{\upsilon}_q,\check{\bm\gamma}).\label{eq21}
\end{align}
\par
From equations (\ref{eq8b},\ref{eq16a}-\ref{eq21}),
we obtain the result of Corollary 1.
\end{IEEEproof}
\medskip
\begin{remark}
Corollary \ref{cor1} gives the performance limitations of some different combinations of network constraints when there is only one single unstable pole and one single NMP zero. The results demonstrates that not only, but also these angles between BTA direction or Euclidean coordinates and NMP zero direction may affect on the tracking performance limitations. And, these angles between the the BN-CAD and the transform direction of NMP zero direction also can affect tracking performance limitations. Moreover, when only the quantization noise and the bandwidth are considered for network's constraints, the bandwidth and the angle between the QNPA direction and the transform direction of NMP zero direction can affect tracking performance limitations, but the bandwidth will just affect the direction of each NMP zero.
\end{remark}
\par
From theorem \ref{thm2}, equations (\ref{eq8b}) and (\ref{eq24}), we can insight into the reference noises only affects the part $J_1^*$, while the network constraints only affects the part $J_2^*$ in the performance limitation. In order to understand the influence of the number of poles and zeros on this network coupling performance limitation part, the corollary \ref{cor2} is given as follows.
\begin{corollary}\label{cor2}
Considering the MIMO system, the NCS structural model is demonstrated in Fig. \ref{fig.2}.
The reference signal $\bm r$, the communication path additive noise process $\bm n$ and the quantization noise $\bm q$ are ought to be a zero mean i.i.d. GWN with PSD ${\sigma_r}^2, {\sigma_q}^2$ and ${\sigma_n}^2$. The encoder $A=\lambda$. $G(s)$ is unstable and NMP plant. Then, this tracking performance limitation can be given as follows.
\begin{description}
  \item[(a)] If $z_k~(k = 1, \ldots ,n_z )$ are NMP zeros of $G(s)$, and  $G(s)$ haven't any unstable pole, then
\begin{align*}
J^*=&2\sum\limits_{i = 1}^{{n_z}} {\mathop{\rm Re}} ({z_i})
\Psi_{r}\Psi_{\eta}\cos\angle ({\bm \upsilon_{r}, \check{\bm\eta}_i})
\end{align*}
  \item[(b)] If $z_k~(k = 1, \ldots ,n_z )$ are NMP zeros of $G(s)$, and there is only one unstable pole $p$ in poles of $G(s)$, then
\begin{align*}
J^*=&2\sum\limits_{i = 1}^{{n_z}} {\mathop{\rm Re}} ({z_i})
\Psi_{r}\Psi_{\eta}\cos\angle ({\bm \upsilon_{r}, \check{\bm\eta}_i})\\
&+{2{\mathop{\rm Re}\nolimits}(p)}{\bm \gamma}^H{\bm f}^H_{net}{\bm f}_{net}{\bm \gamma}
\end{align*}
  \item[(c)] If $p_k~(k = 1, \ldots ,n_p )$ are unstable poles of $G(s)$, and $G(s)$ haven't any NMP zero, then
\begin{align*}
J^*=&\sum\limits_{i,j = 1}^{n_p} \frac{{4{\mathop{\rm Re}\nolimits} ({p_i}){\mathop{\rm Re}\nolimits} ({p_j})}}{{{\bar p}_i}+ {p_j}}\bm \gamma _j^H\tilde{B}_{net, Oj}^{-1}\tilde{B}_{net, Oi}^{-H}{\bm \gamma _i}\\
&\times\bm \gamma _i^H\tilde{B}_{net, Ii}^{-H}{\bm f}^H_{net,i}
{\bm f}_{net,j}\tilde{B}_{net, Ij}^{-1}{\bm \gamma _j}
\end{align*}
  \item[(d)] If $p_k~(k = 1, \ldots ,n_p )$ are unstable poles of $G(s)$, and there is only one NMP zero $z$ in zeros of $G(s)$, then
\begin{align*}
J^*=&2{\mathop{\rm Re}} ({z})
\Psi_{r}\Psi_{\eta}\cos\angle ({\bm \upsilon_{r}, \check{\bm\eta}})\\
&+\sum\limits_{i,j = 1}^{n_p} \frac{{4{\mathop{\rm Re}\nolimits} ({p_i}){\mathop{\rm Re}\nolimits} ({p_j})}}{{{\bar p}_i}+ {p_j}}\frac{(p_i+\bar{z})(p_j+\bar{z})}{(p_i-z)(p_j+\bar{z})}\nonumber\\
&\times\bm \gamma _i^H\tilde{B}_{net, Ii}^{-H}{\bm f}^H_{net,i}
{\bm f}_{net,j}\tilde{B}_{net, Ij}^{-1}{\bm \gamma _j}\nonumber\\
&\times\bm \gamma _j^H\tilde{B}_{net, Oj}^{-1}\tilde{B}_{net, Oi}^{-H}{\bm \gamma _i}
\end{align*}
\end{description}
\end{corollary}
\begin{IEEEproof}
The corollary is easy to prove from theorem \ref{thm2}.
\end{IEEEproof}
\begin{remark}
There are some interesting results have been demonstrated by corollary \ref{cor2}. If if $G(s)$ haven't any NMP zero, $J_1^*$ will be equal to zero. This means we can offset the impact of reference noise under means square sense by 2DOF compensator in this case. Similarly, if $G(s)$ haven't any unstable pole, $J_2^*$ will be equal to zero. This means we also can offset the impact of all network constraints under means square sense in this situation by 2DOF compensator.
\end{remark}

\section{SIMULATION STUDIES}\label{se6}
Given the plant
$G=({s-k})/((s+1)(s-p)),$
In consideration of the encoder $A=\lambda, (\lambda\geq1)$.
Obviously, when $k>0$ and $p>0$, $G(s)$ is NMP and unstable as well.
%\par
%\vspace{5pt}{\bf Case 1:} Simulation for Theorem 1.
\par
In this circumstances, the control scheme by Fig. 1 is considered. The limited bandwidth is generated by the LTI filter. Similar to \cite{09Rojas}, the bandwidth model is considered as the low-pass Butterworth filter with order $1$, and $f_{c}$ is cut-off frequency.
\par
\begin{figure}[H]
 \centering
  \includegraphics[width=7.8cm]{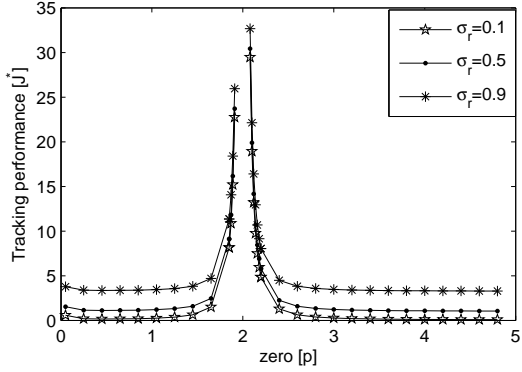}
  \captionsetup{justification=centering}
  \caption{\small{$J^*$ with respect to the plant's unstable poles. $(k=2, f_{c}=2, {\sigma_n}=0.1, b=9, \lambda=2)$}}\label{fig.5}
\end{figure}
\begin{figure}[H]
 \centering
  \includegraphics[width=7.8cm]{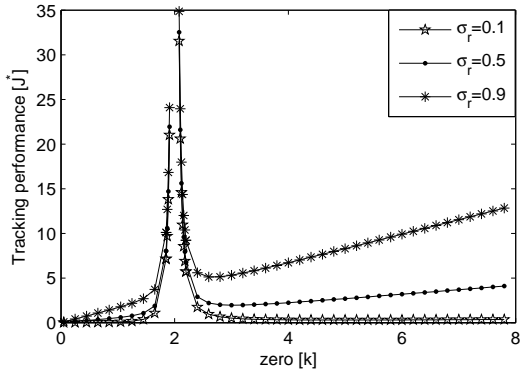}
  \captionsetup{justification=centering}
  \caption{\small{$J^*$ with respect to plant's NMP zero. $(p=2, f_{c}=2, b=8, {\sigma_n}=0.1, \lambda=2)$}}\label{fig.6}
\end{figure}
\begin{figure}[H]
 \centering
  \includegraphics[width=7.8cm]{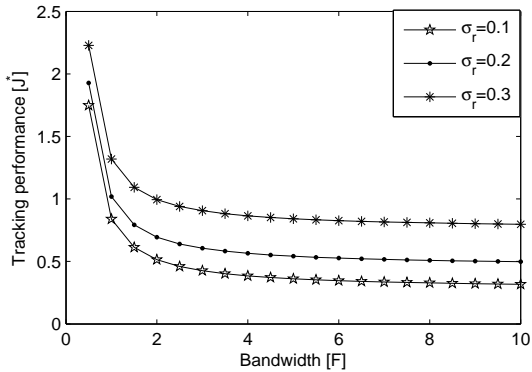}
  \captionsetup{justification=centering}
  \caption{\small{$J^*$ with respect to channel's bandwidth. $(k=3, p=2, {\sigma_n}=0.1, b=8, \lambda=2)$}}\label{fig.7}
\end{figure}

The interaction of the unstable pole's site and the tracking performance limitations is illustrated in Fig. \ref{fig.5}. And accordingly the interaction of the site of the NMP zero and the tracking performance limitations are shown in Fig. \ref{fig.6}. Both the simulations in Fig. \ref{fig.5} and Fig. \ref{fig.6} reveal that the tracking performance limitations will have a tendency to be infinite if $k=2$ or $p=2$ for the reason that pole-zero cancellation happens, and we can conclude what is the same as \cite{10Ding}. As a whole, with the increase of the NMP zero or unstable pole, the degree of influence on the deterioration of performance is greater apart from than the area where the pole-zero cancellation is eliminated. Fig. \ref{fig.7} shows the optimal performance which is constructed by using the channel bandwidth. With a decrease in the usable bandwidth of the communication path, a rise in the optimal performance needed to assure stabilizability is generated accordingly, and similar conclusions can be found in \cite{13Guan}. Fig. \ref{fig.5}, Fig. \ref{fig.6} and Fig. \ref{fig.7} also indicate that the tracking signal will deteriorate the system performance.
%\begin{center}
%\includegraphics[width=8.3cm]{p-sigmar.eps}
%\includegraphics[height=3.1cm,width=7.6cm]{Fig.1.pdf}
%{\footnotesize \centerline{{\bf Fig. 5.} $J^*$ with respect to the plant's unstable poles. $(k=2, f=2, {\sigma_n}=0.1, b=9, %\lambda=2)$}}
%\end{center}
%\begin{center}
%\includegraphics[width=8.3cm]{k-sigmar.eps}
%\includegraphics[height=3.1cm,width=7.6cm]{Fig.1.pdf}
%{\footnotesize \centerline{{\bf Fig. 6.} $J^*$ with respect to plant's NMP zero. $(p=2, f=2, b=8, {\sigma_n}=0.1, \lambda=2.)$}}
%\end{center}

%\begin{center}
%\includegraphics[width=8.3cm]{F-sigmar.eps}
%\includegraphics[height=3.1cm,width=7.6cm]{Fig.1.pdf}
%{\footnotesize \centerline{{\bf Fig. 7.} $J^*$ with respect to channel's bandwidth. $(k=3, p=2, {\sigma_n}=0.1, b=8, \lambda=2)$}}
%\end{center}
\begin{figure}[H]
 \centering
  \includegraphics[width=7.8cm]{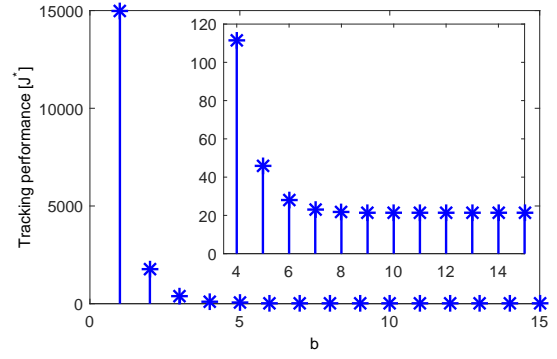}
  \captionsetup{justification=centering}
  \caption{\small{$J^*$ with respect to quantization noise. $(k=2, p=3, f_{c}=2, {\sigma_r}=0.1, \lambda=2)$}}\label{fig.8}
\end{figure}
%\begin{center}
%\includegraphics[width=8.3cm]{b-J.eps}
%\includegraphics[height=3.1cm,width=7.6cm]{Fig.1.pdf}
%{\footnotesize \centerline{{\bf Fig. 8.} $J^*$ with respect to quantization noise. $(k=2, p=3, f=2, {\sigma_r}=0.1, \lambda=2)$}}
%\end{center}
%\begin{figure}[ht]
% \centering
%  \includegraphics[width=8.2cm]{lambda-sigmar.eps}
%  \captionsetup{justification=centering}
%  \caption{\small{$J^*$ with respect to encoder-decoder and reference noise. $(k=2, p=3, f_{c}=2, b=9)$}}\label{fig.9}
%\end{figure}
\begin{figure}[H]
 \centering
  \includegraphics[width=7.8cm]{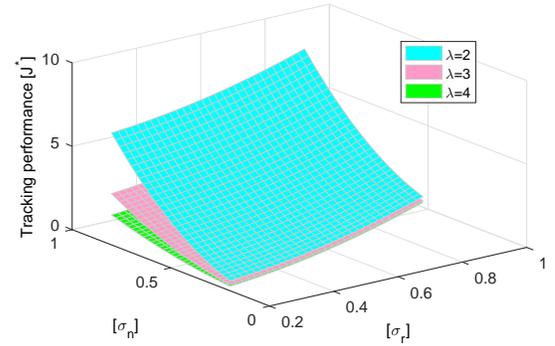}
  \captionsetup{justification=centering}
  \caption{\small{$J^*$ with respect to communication noise and reference noise. $(k=2, p=3, f_{c}=2, b=9)$}}\label{fig.10}
\end{figure}
\begin{figure}[H]
 \centering
  \includegraphics[width=7.8cm]{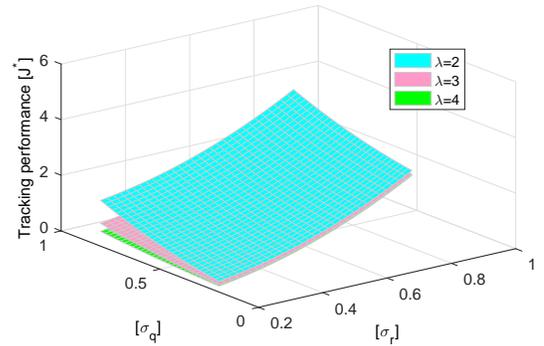}
  \captionsetup{justification=centering}
  \caption{\small{$J^*$ with respect to quantization noise and reference noise. $(k=2, p=3, f_{c}=2, b=9)$}}\label{fig.11}
\end{figure}
In Fig. \ref{fig.8}, it is possible to examine the tracking limitations subjected to quantisation error. For the quantiser level number $b$ consider a range between $1$ and $15$. From Fig. \ref{fig.8}, we can then venture that a 128-bits quantiser will almost recover the infimal tracking performance without noticeable degradation from the quantisation error. It will likewise be observed in the case when it comes to applying a 16-bit quantiser, and consequently, it is needed to prepare to conclude it nearly a six times of the infimal tracking performance for criterion of stabilizability. Fig. \ref{fig.10} and Fig. \ref{fig.11} show the effects of the quantization noise and reference signal, the channel noise and reference signal. Two facts can be appreciated from Fig. \ref{fig.10} and Fig. \ref{fig.11}. Firstly, the reference signal will degrade system performance. The reference noise, channel noise and quantization noise will degrade the tracking performance. Secondly, improving coding level can improve system performance.

%\begin{center}
%\includegraphics[width=8.3cm]{lambda-sigmar.eps}
%\includegraphics[height=3.1cm,width=7.6cm]{Fig.1.pdf}
%{\footnotesize \centerline{{\bf Fig. 9.} $J^*$ with respect to encoder-decoder. $(k=2, p=3, f=2, b=9)$}}
%\end{center}

%\begin{center}
%\includegraphics[width=8.3cm]{b-SNR.eps}
%\includegraphics[height=3.1cm,width=7.6cm]{Fig.1.pdf}
%{\footnotesize \centerline{{\bf Fig. 11.} The relationship between admissible channel SNR and quantization noise. $(k=2, p=3, %\sigma_n=0.2, \mathcal{P}=100, \lambda=2$)}}
%\end{center}

\section{CONCLUSIONS}\label{se7}

In this paper, the tracking performance limitations is  investigated in an AWN channel with bandwidth restriction while considering both reference noise and encode-decode. And furthermore, we apply the ISE criterion to measure the tracking performance. The explicit representations of tracking performance limitation is obtained by the use of $H_2$ optimization technology. It is clearly that the performance limitations for an optimal tracking problem is dependent on the NMP zeros as well as the unstable poles and their directions. What's more, the allocations of the power of reference noises and network communications (quantization noise, communication bandwidth, encode-decode) also have certain effects on the optimal tracking capability. Finally, a typical example shows the effectiveness of our results. Continuous-time systems over communication channels are focused on in this paper. It is delightful that we can develop a discrete-time counterpart to our outcomes accordingly and straightforwardly.\par
In the future, performance limitation allocation may be considered as a promising research direction. This will be analyzed progressively for the time to come. And, it is also worth exploring on performance limitations of distributed NCSs. Additionally, revealing the way the tracking performance limitations are influenced by network-induced constraints, such as packet-dropouts and network delay, is also a meaningful and important project. Our findings are enlightening and guiding in the field of the design of control systems and communication network.

\bibliographystyle{IEEEtran}      %IEEEtran Îª¸ø¶¨Ä£°å¸ñÊ½¶¨ÒåÎÄ¼þÃû
\bibliography{IEEEabrv}

%\appendices
 %\begin{appendices}
 %     \section{Proof of Theorem 1}
 % \end{appendices}
%\section{Proof of Theorem 1}
%\setcounter{section}{0}
%\renewcommand\thesection{\arabic{section}.1}

%\renewcommand\theequation{A.}
%\setcounter{subsection}{0}

%      \section{ 2 }
 %     some text in Appendix B
%\section*{Acknowledgment}

\end{document}